\documentclass[aps,preprint]{revtex4}

\usepackage{graphicx}

\usepackage{epsfig}

\newcommand{\be}{\begin{equation}}
\newcommand{\ee}{\end{equation}}
\newcommand{\nn}{\mbox{} \nonumber \\ \mbox{} }
\newcommand{\ba}{\begin{eqnarray}}
\newcommand{\ea}{\end{eqnarray}}
\newcommand{\om}{\omega}
\newcommand{\Alfven}{ Alfv\'{e}n }

\newcommand{\curl}{{\rm curl\, }}
\newcommand{\E}{{\bf E}}
\newcommand{\B}{{\bf B}}
\newcommand{\J}{{\bf J}}
\renewcommand{\v}{{\bf v}}

\renewcommand{\div}{{\rm \,div\,}}

\newcommand\etal{\textit{et al.\ }}
\newcommand\eg{\textit{e.g.\ }}
\newcommand\cf{\textit{cf.\ }}

\newcommand{\Bf}{{magnetic field\,}}
\newcommand{\Bfs}{{magnetic fields\,}}
\newcommand{\Ef}{{electric  field\,}}
\newcommand{\Efs}{{electric fields\,}}
\newcommand{\CR}{{cosmic ray\,}}
\newcommand{\CRs}{{cosmic rays\,}}

\newcommand{\mnras}{MNRAS}
\newcommand{\apjl}{ApJ Lett.}
\newcommand{\aap}{AAP}
\newcommand{\araa}{ARAA}
\newcommand{\physrep}{Phys. Rep.}

\begin{document}

\title{Inductive acceleration of UHECRs in sheared relativistic jets}
\author{Maxim Lyutikov}
\affiliation{Department of Physics, Purdue University, 525 Northwestern Avenue
West Lafayette, IN
47907-2036 }
\email{lyutikov@purdue.edu}
\author{Rachid Ouyed}
\affiliation{Department of Physics and Astronomy, University of Calgary,
       2500 University Drive NW, Calgary, Alberta, T2N 1N4 Canada}

\begin{abstract}
Relativistic outflows carrying large scale \Bfs have large 
inductive potential and may  accelerate protons to ultra high energies.
We discuss a novel scheme of Ultra-High Energy Cosmic
Ray (UHECR)  acceleration  due to drifts in 
 magnetized, cylindrically collimated, sheared jets of powerful active galaxies
(with jet luminosity $\geq 10^{46}$ erg s$^{-1}$). We point out that a
 positively charged particle carried by such a plasma is in an unstable  
equilibrium if  ${\bf B} \cdot \nabla \times {\bf v}< 0$, 
so that kinetic drift along the velocity shear would lead
to fast, {\it regular} energy gain.  This can be achieved in an  axially 
 inhomogeneous jet through gradient drift induced by propagation of inertial
\Alfven waves along the jet.
We show that if a seed of pre-accelerated particles with energy below
the ankle $\leq 10^{18}$ eV  is present,  these particles
can be boosted to energies above $ 10^{19}$ eV.
A key feature of the mechanism is that the 
highest  rigidity  (ratio of energy to charge) particles  are 
accelerated most efficiently implying
 the dominance of light nuclei for energies above the ankle in our model: from a mixed population
of pre-accelerated particle the drift mechanism  picks up and boosts protons
preferably.  In addition, 
after  a particle
traversed
large fraction of the available potential, its  Larmor radius becomes
of the order of the jet thickness.
In this case, {\it the   maximum  possible acceleration rate of inverse relativistic gyro-frequency
is achieved} and a particle finally
 become unconfined and leave the jet.
 The power-law spectrum of the resulting UHE particles  flattens
 with time and asymptotically  may become $\propto E^{-2}$. The real injection spectrum depends on the distribution of pre-accelerated particles inside a jet and, in case of contribution from many sources, on the distribution of total potential drop.

We also point out that  astrophysical schemes based on DC-type acceleration
(by electric
field parallel to or larger than  magnetic field)
cannot have potentials larger than $\sim 10^{15}$ Volts
and thus fell short by many orders of magnitude to produce UHECRs.

\end{abstract}
\maketitle

\section{Introduction}

Acceleration of Ultra High Energy Cosmic Rays (UHECRs), with energies
reaching  $ 3 \times 10^{20}$ eV, remains
one of the main challenges of modern physics and astrophysics
 \cite{bland00,olinto00,ostrow02a,torres04,parizot05}. 
UHECRs can be either accelerated by  ''astrophysical Zevatrons'' 
(or bottom-up scenario) or a be a by-product of decay of massive  exotic particles
(top-down scenario). 

The most commonly discussed astrophysical  mechanism of acceleration of UHECRs 
is
 diffuse shock acceleration  by relativistic shocks \cite{kirk99,baring04} where
 small scale \Bf,  amplified typically
 to near-equipartition (see \S \ref{cites}), are required.
In addition, it is necessary  that  scattering rate is close
to the maximum (Bohm) rate in order to overcome advection of 
particles downstream of a shock.
It is far from clear whether such rates are achievable \cite{niemiec04}. 

Commonly, it is believed that 
low energy \CRs, with energies approximately 
below the ''ankle'' at $\sim  10^{18} $ eV,
are of Galactic origin (GCR), while those above the ''ankle'' 
are extragalactic. [Note, that exact location of transition between Galactic and extragalactic \CRs 
is not clear and may instead be associated with  the ''second knee'', \cite{wick04}.]
Since spectra below and above the ''ankle''  are different,
different acceleration mechanisms  should operate \cite[see][for an alternative
holistic scenario]{allard05}.
In this paper we propose 
new mechanism that operates at highest energies and thus is responsible
for extragalactic \CRs (EGCR): 
particle acceleration by inductive electric  fields in magnetized,
relativistic sheared  outflows. 

Inductive \Efs and \CR acceleration
 has been mentioned by many authors \citep{bland00,levinson00,
blasi00, arons03}, but no concrete acceleration mechanism
has been proposed (with a possible exception of \cite{bell92}, see below).
It was only pointed out that there is a large electric potential and it was {\it assumed}
that particle somehow manage to tap it. 
Since inductive \Efs are orthogonal to \Bf and particles cannot move freely along them,
it is not obvious how to achieve energy gain.
Another acceleration scheme is acceleration in DC electric field
aligned with \Bfs  (double layers), but these, as we demonstrate in appendix \ref{double},
cannot account for UHECRs.

\subsection{Inductive electric fields}

As mentioned above, inductive \Efs are not easily accessible for acceleration
since particle need to move across \Bf, a processes prohibited under 
ideal Magneto-Hydrodynamics (MHD) approximation.  On the other hand, 
kinetic drifts may result in   regular motion  across \Bf and along \Ef
leading to  {\it regular energy gain} as compared
to the stochastic, Fermi-type schemes.
 Since the  drift velocity  increases with particle energy, {\it the rate of energy gain 
will also  increases with particle energy}. Thus, the highest energy
particles  will be accelerated most efficiently. This means that 
from a pre-accelerated population, the mechanism proposed
here will pick up particles with highest
energy and will boost them to even  higher values.
In addition, when a particle has crossed a considerable fraction
of the total available potential,
it gyro-radius becomes comparable with flow scale (\S \ref{cites}).
In this case, drift approximation brakes down. As a result,
as  long as  the  particle remains inside a jet,
{\it the  acceleration rate reaches the
theoretical maximum of an inverse gyro-frequency  (\S \ref{rL})}.
Thus, the most efficient acceleration occurs right before the particle leaves the jet.
One may say that
 in case of inductive acceleration, {\it becoming unbound is beneficial
 to acceleration}, contrary to the case of stochastic acceleration when
 for unbound particles acceleration ceases.

Typically, the direction of drift is along the normal to the \Bf and to the
direction of the force that induces a drift.
 In sheared cylindrical jet with toroidal \Bf
electric field is in radial direction, so that 
 in order to gain energy particle should  experience 
radial drift. It is then required that there should be a force along the
axis. Such force may arise if   a jet is axially  inhomogeneous 
resulting in gradient drift due to changing magnetic pressure.
Several types of inhomogeneity can occur in astrophysical jets.
Numerical simulations of hydromagnetic jets show complicated axial structure
with regions of compression and rarefaction \cite{vanPut96,ouyed97,komisarov99}.
Besides shocks, these simulations show  prominent subsonic density and 
\Bf variations inside jets.  Subsonic  variations propagate roughly with the jet flow 
and  induce axial variations of \Bf   leading to radial  drifts.
An important property of the mechanism proposed here is  that for a particle at 
rest with respect
to the bulk flow energy gain (or loss) is {\it  independent }  of the direction of
the drift (\eg away or towards the axis; see \S \ref{6}). On the other hand, 
if an axial variation is periodic, \eg a wave, 
changing the direction of the drift would lead to alternative energy gains and losses.
If in one wave period a high energy particle  acquires a large enough  energy so that
its gyro-radius  becomes comparable to jet scale, it  will leave the jet.

\subsection{Electric drifts and particle acceleration}

To separate effects of electric  drift in crossed electric and \Bfs 
(which does not lead to energy gain) and particle acceleration, 
we should consider particle dynamics in a local
plasma rest frame, where electric field vanishes at the origin. Since the 
velocity field is likely to be sheared (\eg higher velocity near the axis of a jet), 
there will be non-vanishing electric field away from the origin. 
Value of the electric field at a position of a particle
will be smaller than inductive field in the laboratory frame, approximately, by a ratio
of Larmor radius to shear scale. Typically, both  shear velocity and  \Ef would be
linear function near the origin, so that {\it  electric potential is
 a quadratic function of distance from  the origin of the local plasma rest frame}.
Depending on the sign of the coefficient in front of the quadratic term,
electric potential has either a minimum or maximum near the origin.
This depends on the sign of the charge and sign 
of the product $  \B \cdot \nabla \times {\bf v}$. In \S \ref{motion} we show that
{\it in sheared flows with  $ \B \cdot \nabla \times {\bf v}< 0$
protons are near the maximum of electric potential}, so that drift motion
away from the  origin ({\it along or against}  the direction of the shear), will lead to 
energy gain. 

Our model bears some resemblance to  Bell model \cite{bell92}  of \CR acceleration 
at pulsar wind nebular, which relies on 
large (unreasonably large, in our view) \Bf gradients near the axis of the expanding 
flow in order  to ''pull-in'' pre-accelerated charges against weakly relativistic expansion of 
plerion.  The proposed mechanisms  also has both  similarities and    differences from the
shock drift acceleration at non-parallel shocks \cite{webb83}. 
A shock can be considered as an extreme case of field inhomogeneity
(a delta-function), so that drift velocity is a delta function
 \citep{Jokipi82}. This can similarly produce energy gains due to displacements of 
particle along inductive electric field associated with
shock motion.
The idea that particle energy can be limited by ``lateral'' escape, is 
also reminiscent of Eichler  model   \cite{eichler81} 
of particle acceleration at the Earth's bow shock. 
The difference is that in Eichler's model the sideways motion 
of a CR was due to cross-field diffusion, while in our case it is a gradient drift.

Our acceleration scheme is also very different from those previously considered 
 in sheared flows due to 
second-order Fermi process \cite{Berezhko81,Jokipi90,ostrow02a,rieger04}.
In the references cited above particle energetization occurs due 
elastic scattering by \Bf inhomogeneities in the flow rest frame.
Acceleration is stochastic, while  acceleration time scales
are typically very long, scaling as velocity shear squared, $V'^2$,   \cite{Jokipi90},
so that extremely narrow layers are required.
Contrary to this, in our scheme  acceleration   is regular and scales as $V'$.

Let us also point out that any acceleration scheme based on DC-type acceleration,
where there is large electric  field along \Bf or where \Ef becomes
larger than  \Bf, cannot accelerate particles to energies above
$10^{15} $ eV, appendix \ref{double}. This excludes, for example, magnetic reconnection
\cite{lesch02,kcld04}.

In this paper we start in Section 2 by briefly describing on general
 basis the properties of UHECR acceleration sites. The constraints
on source energetics and size are discussed within the context of
cylindrical AGN jets. In Section 3 and 4 we study the evolution
of particles trapped in sheared flows and in the presence
of Alfv\'en waves. The details of the acceleration mechanisms
at play are described in details in Section 5. The corresponding
 particle spectrum is derived in Section 6 before considering
the astrophysical implications in Section 7. A discussion
is presented in Section 8 and a brief conclusion in Section 9.

\section{AGN jets as Possible sites of UHECR acceleration}
\label{cites}

\subsection{Luminosity - Potential relation}

In  bottom-up acceleration schemes \cite{torres04,bgg06} any 
source that accelerates a particle to a particular energy
should have a total electric potential drop of the same order 
(it is unlikely that a particle completes many circuits with much  smaller potential drops). 
On very basic grounds, we expect that in any astrophysical system  electric field are no 
larger than \Bf. [Localized gaps, regions with either parallel electric field or
total electric field larger than magnetic, may exist but the typical
potential drop in them is small (for purposes of UHECR acceleration) and
is limited by lepto-photonic  pair production, \cite{arons79}.]
This has two consequences. First, regardless of the acceleration mechanism,
the highest possible acceleration rate
is inverse of the relativistic  gyro-frequency, $\sim \gamma/ Z \om_B$ where $\om_B = e B/mc$
is cyclotron frequency, $Z $ is a charge of a particle in terms of elementary charge $e$,
$m$ is a mass of a particle.
Secondly, we can place very general 
constraints on possible accelerating sites \cite[see also][]{lovelace76}.
If we parameterize $E = \beta_0 B$ where $E$ is the \Ef and $B$ is the \Bf 
($\beta _0$ is the ratio of typical electric to \Bfs) and 
 if an accelerating region has a size $R$  the available
potential is $\Phi \sim  R E \sim \beta_0 R B$ and the energy of particle
crossing that potential would be
\be
 {\cal E} =  Z e \beta_0 B R
 \label{1}
 \ee
 For $\beta_0 \sim 1$ this is also the condition 
that the particle's gyro-radius is smaller than $R$ \cite{Hillas}. 
Thus, for $\beta _0 \sim 1$, a system will be able to accelerate particles to the
maximum energy that it can confine.
We would like to stress that the conditions on acceleration of UHECRs
(\ref{1}) are different from confinement condition $r_L <  R$, where $r_L$ is the Larmor radius.
The two conditions are different by a factor $\beta_0$:
 for  $\beta _0 \ll 1$ a system can confine
particles with energies much larger than those to which it can accelerate.
 
Since $B R$ is a measure of the total current in the system, Eq. (\ref{1}) can be rewritten
as  (assuming $\beta_0 \sim 1$)
\be
 \Phi = \beta_0 {2  I \over  c}
  \label{1a}
 \ee
This determines the maximum electric potential and corresponding 
energy ${\cal E} =  Z e  \Phi$ to which a system carrying a current $I$ can accelerate. 

Unless the electric field is strictly along the \Bf (\eg in double layers, 
see appendix \ref{double}), 
there is Poynting flux in the system. Since  Poynting flux cannot be larger than 
 the total luminosity of a given source, assuming a spherical expansion  we find
\be
L = 4 \pi R^2 {1+ \sigma \over \sigma }
{ E B \over 4 \pi}   c \sim  {1+ \sigma \over \sigma }
R^2 B^2 \beta_0 c =   {1+ \sigma \over \sigma } {1 \over \beta_0}  \Phi^2 c
\label{L}\\
\ee
where $\sigma$ is the ratio of Poynting to particle flux \cite{KC84}

Thus, assuming $\beta_0 \sim 1$,
\ba &&
  \Phi \leq  \sqrt{  \sigma \over 1+ \sigma } \sqrt{ 4 \pi  L \over c} =
  4 \times 10^{20} {\rm V} L_{46}^{1/2}
  \nn &&
  I \leq  \sqrt{ \sigma \over 1+ \sigma } \sqrt{ L c \over  4 \pi} =
  3.6 \times 10^{25} {\rm A}  L_{46}^{1/2}
  \nn &&
  L = {1+ \sigma \over \sigma } I \Phi
  \label{phi}
  \ea
  where we introduced coefficients of the order of unity
 so  that in the limit $\sigma \rightarrow \infty$
 the system impedance is that of free
  space, $4 \pi/c$; $L_{46}=L/10^{46} $ erg/s and for numerical estimate we assumed $\sigma =1$. 
  For each particular system these relations will be modified by 
  a numerical factor which depends on geometric properties of the system.
The case of cylindrically collimated AGN jets is considered in \S \ref{cyljet}.

Eq. (\ref{phi}) relates the maximum available 
 potential to the luminosity of the source. 
Assuming that UHECRs are protons, we can select possible candidates
for UHECRs acceleration based on their luminosity: to  achieve $3 \times 10^{20}$ eV we need $L\geq 10^{46} $ erg/s. 
We can  immediately  {\it exclude acceleration of UHECRs in low power
AGNs  (\eg Cen A and  M87), 
low power BL Lacs and starburst galaxies (\eg M82 \& NGC 253)} (in case of proton component),
 see also \S \ref{Astr}.
 This limits the possibilities to more distant high power AGNs like 
 higher power FR I, 
FR II radiogalaxies, radio-loud quasars  and GRBs.
[In appendix \ref{grbs} we argue against GRB shocks as source of UHECRs.]
 Luminosity constraint (\ref{phi}) also excludes, in our view, holistic scenario, \cite{allard05},
where the same acceleration mechanism operates in all galaxies, including our own.
Only ''special'' galaxies, satisfying condition  (\ref{phi}) are able to accelerate UHECRs.
Note, that in order to maximize potential at a given luminosity,
it is required that $\sigma \geq 1$, but not necessarily $\gg 1$.

Another constraint that acceleration sites should satisfy is that
radiative losses should not degrade particle energy.
 We can derive very general constraints
on possible location of \CR acceleration just by balancing 
most efficient acceleration, by $E\sim B$, and 
radiative losses \cite[see also][]{medved03}. Typically, radiative losses scale as
$\propto \gamma^2 u$  where $\gamma$ is the Lorentz factor of a \CR
and $u$ is energy density of the medium that leads to radiative losses.
For synchrotron losses $ u=u_B=B^2/8 \pi$, for IC losses
$ u=u_{IC}$ is the energy density of soft photons.
If we normalize total energy density to energy density of \Bf
$u= \zeta u_B$, $\zeta > 1$, 
and 
equate the maximal rate of  energy gain
$d {\cal E} /dt =  Z e c E \sim Z e c B $ to radiative losses
$d {\cal E} /dt \sim  {Z^4 e^2 \over c} \om_B^2 ( {\cal E} / mc^2)^2$
we find
\be
{\cal E}  = mc^2 \sqrt{ c  \over \zeta r_c Z^3 \om_B}  
\label{2}
\ee
where $r_c = e^2/m c^2$. 
This
 may be considered as a second constraint on the site of particle acceleration.
Resolving eqns (\ref{1}) and  (\ref{2}) 
with respect to $B$ and $R$ we find
\ba &&
B < { m^2 c^4 \over \zeta  Z^3 e^3}  \left( { m c^2 \over {\cal E} } \right)^2 \Gamma^3
= 2 \, \Gamma^3 \, {\rm G} \,  \left( { {\cal E} \over 100 EeV} \right)^{-2}  
\left( {1 \over Z} \right)^3
\nn &&
R >   {Z^2 e^2 \zeta  \over m c^2}  \left( { {\cal E} \over m c^2} \right)^3 { 1 \over \Gamma^2} =
 10^{17} \, { 1 \over \Gamma^2} \, {\rm cm} \,
 \left( { {\cal E} \over 100 EeV} \right)^3  \left( Z \right)^2
 \label{B}
\ea
 Where we also allowed a possibility that plasma is expanding with bulk Lorentz factor
 $ \Gamma$, so that \Bf in the plasma rest-frame  is $B/\Gamma$ and
 typical scale is $R/\Gamma$, and assumed $\zeta=1$; $EeV= 10^{18}$ eV. 

Relations (\ref{B}) show that higher energy \CRs are better accelerated at large distances.
 AGN jets, which propagate
to more than  100 kpc distances present an interesting possibility. Note, that as long as the jet remains relativistic, the total inductive potential is approximately  conserved, 
so one can  ``wait'' a long time for a particle to get accelerated without worrying about radiative loses.  Thus, UHECRs can be accelerated  inside the jet at distances from a fraction
of  a parsec (Eq. (\ref{B})) to hundreds of kpc, as long as the jet remains relativistic
and sustains a large inductive potential.


\subsection{Cylindrical AGN jets}
\label{cyljet}

As an illustration of the general principles discussed
above let us consider
 force-free
 cylindrically collimated jets.
A
 possible structure of such a jet is a relativistic sheared diffuse pinch
\ba && 
B_\phi = {r/R_c \over 1+ (r/R_c)^2} B_0
\nn && E_r = \sqrt{1-1/\Gamma(r)^2} B_\phi
\nn && B_z= {1 \over 1+ (r/R_c)^2} B_0
\ea
where  $R_c$ is a radius of a current-carrying core.
System remains in force-balance for
 arbitrary $\Gamma(r)$ \citep{lpg04}.
 Alternatively, instead of $B_z$ particle pressure may ensure force-balance.
 
 For illustration purposes let us choose a Lorentz factor with a profile,  
 \be
 \Gamma(r)= \sqrt{1+{\Gamma_c^2 -1 \over  1+ (r/R_c)^2}}
 \ee
 where $\Gamma_c=1/\sqrt{1-\beta_c^2}$ is the Lorentz factor of the fast core and we assumed that
Lorentz factor changes on the scale of the core radius.
 Then the total potential
 is 
 \be
 \Phi_{tot} = B_0 R_c {\rm arctanh}(\beta_c) 
 \approx {B_0 R_c \over 2} \ln {2 \over 1- \beta_c},
 \ee 
 while total the Poynting flux is
 \be
 L = {B_0^2 R_c^2 c \over 4} 
  \left( {1+\beta_c^2 \over \beta_c^2} {\rm arctanh} (\beta_c) - {1 \over \beta_c } \right) \approx {B_0^2 R_c^2 c \over 4} \ln {2 \over 1- \beta_c}
  \ee
 where the last equalities assume $\beta_c \rightarrow 1$.
 Thus
 \be 
  \Phi_{tot} =\sqrt{L\over c} \sqrt{  \ln {2 \over 1- \beta_c}} \approx
 \sqrt{L\over c}  \sqrt{2 \ln 2 \Gamma_c}\\
  \ee



If a particle start on the axis with $\gamma=\gamma_c$, and drifts in a radial
direction, its
 Lorentz factor $\gamma$ changes with radius as
\be
\gamma = \gamma_c + { \om_{B,0}  R_c \over c}
\left( {\rm arctanh} \beta_c - {\rm arctanh} { \beta_c \over 1+(r/ R_c  \Gamma_c)^2} 
\right)
\ee
so that the particle gains large fraction of the potential after reaching
$r \sim  R_c  \Gamma_c$. At this point its Larmor radius $r_L$ would become
\be
{r_L \over R_c  \Gamma_c} \sim  {\rm arctanh} \beta_c  \geq 1
\ee
This illustrates an important point:  after a particle has crossed a large fraction of the
total available potential, its Larmor radius  becomes comparable to
the flow scale, so that it will decouple from the flow.


\section{Potential energy of a charge in sheared flow}
\label{motion}

Our model of UHECR acceleration relies on the observation that in a
transversely sheared flow one sign of charges is located at a maximum of electric
potential, as we describe in this section. Consider 
 sheared flow carrying \Bf. At each point there is electric field $\E = - {\bf v} \times \B/c$, so that the 
electric potential is determined by
\be
\Delta \Phi = { 1 \over c}  \nabla \cdot \left({\bf v} \times \B  \right)
=   { 1  \over c}  \left( \B \cdot \left( \nabla \times   {\bf v} \right)
-{\bf v} \cdot \left( \nabla \times  \B \right)
\right)
\label{Phii}
\ee
In a local rest frame, the second term in Eq. (\ref{Phii})
vanishes at the position of a particle and is generally sub-dominant to the first term  for $v \ll c$
 \footnote{Generally, $\nabla \times  \B \neq 0$ due to advective current $\sim  ( \nabla \cdot \E ){\bf v}$. In other words, $B /sqrt{1-(v/c)^2} $ is constant, but not the value of \Bf itself.}. 
Then $\nabla \times  \B=0$ and we find \footnote{Keeping the ${\bf v} \cdot \left( \nabla \times  \B \right)
$  term  in Eq. (\ref{Phii}) and assuming force-free balance $ \E\, {\rm div} \E + {\bf j} \times \B=0$, we find ${\bf j}= {\rm div} \E \,{\bf v}$, which gives 
$\Delta \Phi = {   \left( \B \cdot \nabla \times {\bf v} \right)\over c (1 +(v/c)^2)}$.}
\be
\Delta \Phi = { 1 \over c} \left( \B \cdot \nabla \times {\bf v} \right)
\label{Phi}
\ee
We have arrived at an important result:
{\it  depending on the sign of the quantity 
$ \left( \B \cdot \nabla \times {\bf v} \right)$ (which is a scalar) charges  of 
one  sign are  near potential minimum, while those with the
opposite sign are near  potential maximum. }
Since electric field is perpendicular both to velocity and magnetic field, locally, the
electric potential is a function of only one coordinate along this direction.
For $ \left( \B \cdot \nabla \times {\bf v} \right) <0 $ (we will call this case negative shear)
ions are near potential maximum. 

The procedure outlined above to calculate electric potential
is beyond the limits  of applicability of {\it non-relativistic}
MHD, which assumes quasi-neutrality and thus neglects the dynamical effects associated with the 
potential (\ref{Phi}). Thus, even in the low frequency regime with non-relativistic velocities, 
a conventional realm of MHD theory, one should use at least two fluid approach and 
also must retain both charge density as well as
displacement current in Maxwell equations.

We can arrive at a more general result using a relativistic
MHD formulation. An invariant charge density is given by
\be
\rho = j^{\mu} u_{\mu} =   {1\over 4 \pi} F^{\mu \nu} _{, \nu}  u_{\mu}
\ee
where $ j^{\mu}$ is four-current, $ u_{\mu}$ is fluid four-velocity
and $F^{\mu \nu}$ is the electromagnetic tensor and Greek indexes run $0,1,2,3$
and we set speed of light to unity here.
In the framework of  relativistic MHD, when stress-energy tensor
is diagonalizable, the electromagnetic tensor can be written as 
$F^{\mu \nu} = \epsilon ^ {\mu \nu \alpha \beta} B_ \alpha u_\beta$,
where $ B_ \alpha$ is a four-vector of magnetic field and $\epsilon ^ {\mu \nu \alpha \beta}$ is a unit antisymmetric tensor.
Charge density becomes
\be
\rho = {1\over 4 \pi}  \epsilon ^ {\mu \nu \alpha \beta}  u_{\mu} 
\left(  B_ {  \alpha,\beta} u_\beta +
B_ \alpha  u_{\beta, \nu} \right)
\ee
If the \Bf is stationary and homogeneous, $ B_ {  \alpha,\beta}=0$,
then we arrive at
\be
\rho = {1\over 4 \pi} F^{\mu \nu}  u_{\mu, \nu}
\ee
which generalizes Eq. (\ref{Phi}) to an invariant form.


Under ideal fluid approximation particles cannot move 
across \Bf lines, so that they cannot
``sample'' the electric potential  (\ref{Phi}). 
On the other hand, kinetic effects, 
like drift motions, may lead to regular radial displacement  along the shear and thus along
electric field.  In this case
 one sign of charge will be gaining energy, while
another sign will be losing energy. This is independent on whether
the drift is along the  shear  or counter to the shear direction
and thus is independent on the sign of the \Bf gradient that induces the shear.

For low energy particles, with  small  but finite Larmor radii,
  variations of potential energy need to be taken into account
for a correct description of the plasma distribution
function in magnetized  flows with transverse shear
\cite{ganguli88,cai90,ganguli95}. 
 [Since the distribution
 function should depend  on constants of motion, in presence
 of a velocity shear these constants are different from the case of no shear.]
Kinetic motion allows particle to probe regions with different potential and lead to
redistribution of positive and negative changes in an effort to screen 
the shear-induced electric field. These effects are absent if background plasma is cold.
On the other hand, for a small population of high energy particles with large Larmor
radii, electrostatic potential (\ref{Phi}) may become important.

\section{Acceleration mechanism in a nutshell}

Our key point is that an electrostatic cross-field potential 
(\ref{Phi}) may be used for particle acceleration. In this Section we briefly outline the 
principal astrophysical and plasma physical issues related to this acceleration mechanism.
In appendix \ref{waves} and \S \ref{6} we elaborate in
more details the qualitative statements made in this Section.

There is a consensus that relativistic outflows, and  AGN jets in
particular, are accelerated to relativistic speeds and collimated by
large scale \Bfs threading accretion disk and central black hole.
This is  based mostly on theoretical and numerical models of jet launching
from black hole-- accretion disk systems
  \cite{bla77,bp82,begelman84,Gammie}. In addition,
  polarization properties of 
 parsec scale  AGNs jets \citep{lpg04} and, possibly, GRBs
\citep{lpb03} are also consistent with the presence of large scale
\Bf. Energetically, \Bfs may carry a large 
fraction of jet luminosity \citep{bla02,lb03}, so that
 the parameter $\sigma$ introduced above is not too small, $\sigma \geq 1$.

As the jet expands, 
the ratio of toroidal to poloidal \Bf increases approximately linearly
with cylindrical radius of a jet, so that at largest scales \Bf is dominated
by a toroidal component $B_\phi$. In addition,
  toroidal \Bf may provide jet collimation, so that asymptotically AGN
jets may also be fully collimated to a cylindrical shape \cite{HeyN03}. 
One also expects that  jets are sheared, so that the 
central spine of the jet is moving with larger velocity than
its periphery \cite{hp91,Chiaberge00,Giroletti04},
so that $v_z = v_z(r)$ ($r$ is the radial coordinate in cylindrical geometry)
is decreasing outward.  In this case,  
 $ \left( \B \cdot \nabla \times {\bf v} \right)=- B_\phi \partial _r v_z$.
 If, in addition, the jet is inhomogeneous along the axis, $B_\phi= B_\phi(z)$, there will be
 a gradient drift  along $ r$, which in case of negative shear would lead to proton 
acceleration.  In another type of jets, with positive shear $ \left( \B \cdot \nabla \times {\bf v} \right)> 0$,
electrons are accelerated while protons 
lose their gyrational energy eventually coming to a rest with respect to the bulk flow.

An important requirement on the proposed  acceleration  scheme is that
the flow be nearly cylindrically collimated. This stems from the fact that
as long as a particle has not crossed the total available potential,
its drift velocity is non-relativistic, so that in conically expanding relativistic flows
 energy gain cannot compete with adiabatic losses. 

\begin{figure}[ht]
\includegraphics[width=0.9\linewidth]{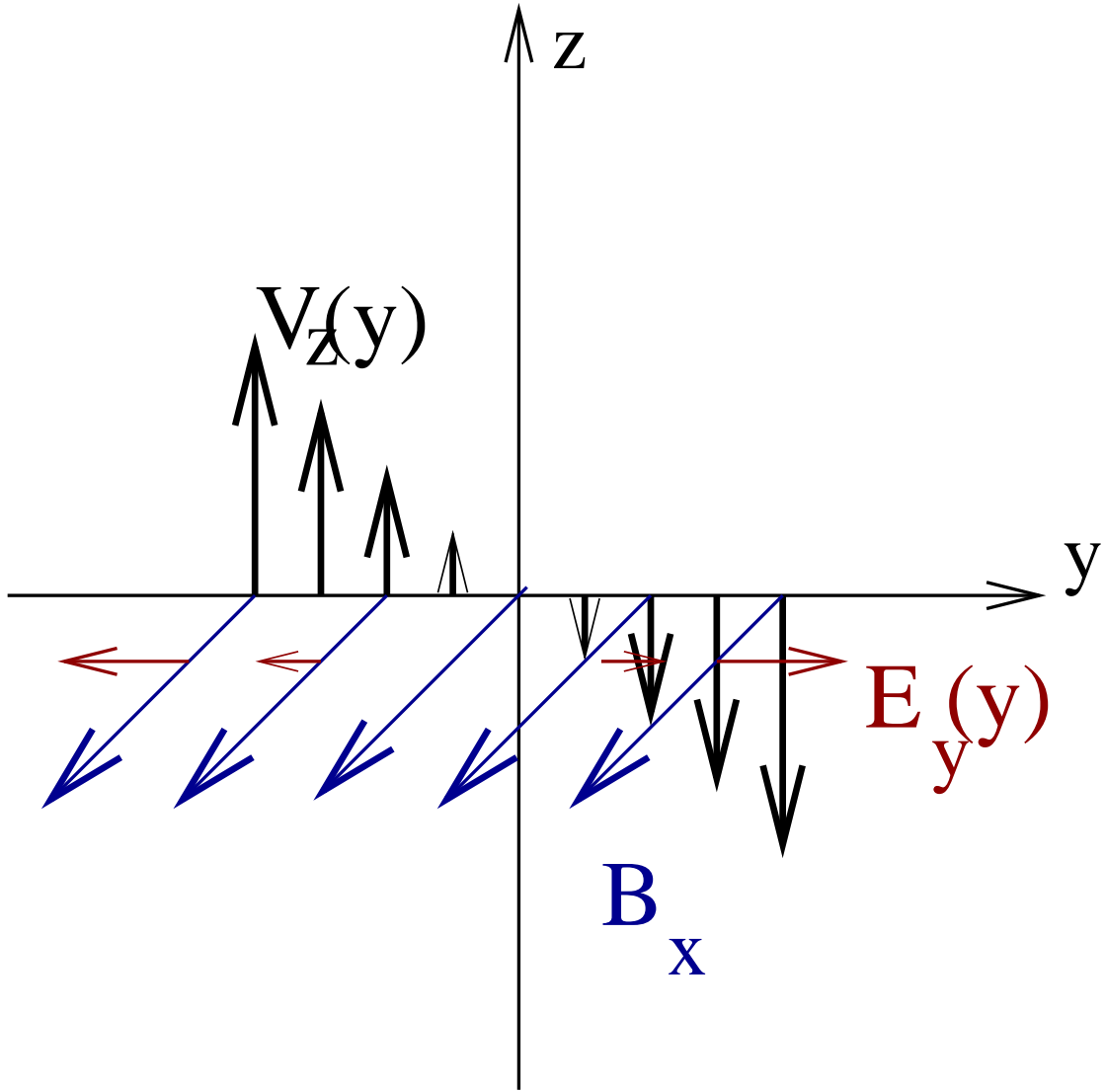}
\caption{Electric field in a sheared flow in a frame where surface the $y=0$
is at rest.  For $\B \cdot \nabla \times {\bf v} < 0$
 the \Ef is directed away from the $y=0$ surface. }
\label{shearCR}
\end{figure}

Next we discuss the salient properties of kinetic particle drift
motion and energy gain in  sheared flows.
Consider a plane-parallel sheared flow in which the \Bf is perpendicular 
to the velocity. For definiteness,  let \Bf be directed along $x$ axis, $B_{x}=B_0$ and 
velocity along $z$ axis, $V=V_{z}(y)$ (see Fig. \ref{shearCR}).
Let us transform to a frame moving with the
fluid at some particular values of $y=y_0$, e.g.  at $y_0=0$.
Expanding velocity near $y=0$, 
$V_{z} = \beta_0 c y/L_V$, where $\beta_0$ is characteristic velocity in terms of speed of light
and
$L_V$ is a characteristic scale. 
Thus, there is electric field 
${\bf E}=- V_{z} B_{x}/c \vec{e}_y =- (V_0 /c) B_0 y/L_V \vec{e}_y=- \beta_0  B_0 y/L_V \vec{e}_y$ and
 electric potential 
\be
\Phi=  \beta_0  B_0 {y^2 \over 2 L_V}
\ee

Next, assume that there is a long wavelength inertial \Alfven wave propagating
along the $z$-direction with a phase speed $V_A$. For $V_A \ll c$, 
 the magnetic perturbation  $\delta B$ 
in the wave is much larger than electric perturbations $\delta E$
by a factor $c/V_A$, so that the wave is nearly magneto-static
(in other words $ \om \ll k_{z} c$). A test particle will experience a drift
in the $x$-direction with magnitude
\be
u_d \sim {\delta B \over B_0} { \gamma c^2 k_{z} \over Z \om_{B,0} } \sim { \gamma c^2 k_{z} 
\over Z \om_{B,0}} 
\ee
where we assumed strong perturbation $\delta B_0 \sim  B$ 
and $\om_{B,0} = e  B_0/mc$.
As a particle drifts along \Ef in the $x$-direction its Lorentz factor 
evolves according to
\be
\partial _ t  \gamma = {Z e E_{y} u_d  \over m c^2} =   \beta_0{ Z \om_{B,0}  u_d^2 t \over c L_V}
\sim \beta_0  { \gamma^2 c^3 k_{z}^2 t \over Z \om_{B,0}  L_V}
\label{q1}
\ee
Thus, energy gain or loss of a test particle depends
on direction of \Bf,  sign of charge (through $\om_B$) and 
direction of velocity vorticity (through sign of $L_V$):
in other words it depends on sign of shear. On the
other hand, it is {\it independent} of the direction
of the drift.

Since drift velocity increases with particle energy, the rate of energy gain
also increases with energy, see Eq. (\ref{q1}). This leads to 
one of the most unusual properties of the proposed acceleration mechanisms:
 {\it  highest energy (or highest rigidity)
 particles are accelerated most efficiently}. In addition,
 at the last stages of acceleration, when particle
Larmor radius becomes of the order of jet scale, particle motion in positive shear flow
becomes unstable even without gradient drift while  {\it  acceleration rate
does reach absolute theoretical maximum of inverses relativistic gyro-frequency }
(\S  \ref{rL}).

In addition, since acceleration rate is {\it inversely}
proportional to charge,  
at a given energy small charge (higher rigidity) particles
are accelerated most efficiently. Thus, from a population 
of pre-accelerated particles with mixed composition, drift mechanism
will pick up particles with smallest charge: protons.
This explains why above the ankle protons start to dominate over heavy
nuclei.

Above is the upshot of the proposed acceleration mechanism.
Next  we elaborate its details.
In appendix \ref{waves} we consider properties of 
electromagnetic waves in transversely sheared cold plasma; this is required
to in order to set up correctly the background electromagnetic fields
for the drift motion of a particle.
In \S  \ref{6} we consider  drift  motion of test particle in
an electromagnetic fields of a low frequency inertial
\Alfven wave propagating in a sheared flow in case of weak shear,
when particle gyro-radius is much smaller than shear scale.
This is done analytically for weak waves and numerically for strong waves.
In \S  \ref{rL} we consider particle motion 
for strong shear, when shear scale is comparable to gyro-scale,
so that drift approximation is inapplicable.
In this case a new type of acceleration mechanism turns on,
which does not require presence of velocity-aligned \Bf gradients
and which leads to theoretically maximal acceleration rate
on inverse of relativistic gyro-frequency.

\section{Details of the acceleration mechanism}
\label{6}

\subsection{Stage 1: Kinetic drift}
\label{drift1}

\subsubsection{Kinetic drift in linear \Alfven waves}
\label{linear}

In this Section we first consider kinetic
motion of a particle in a combined electromagnetic fields
of sheared flow and a low frequency inertial \Alfven wave propagating along 
velocity. For simplicity we concentrate on planar problem and assume a linear 
velocity profile.
Using results of  appendix \ref{waves}, 
electromagnetic fields of the inertial \Alfven are then given by
\ba &&
\delta E_{y}= a B_0 \sin \left ( k_{z} ( z- (V_A -V_{z})t)\right )
\nn &&  
\delta B_{x} =- { a B_0 c\over V_A -V_{z}} \sin \left ( k_{z} ( z- (V_A -V_{z})t)\right )
\label{Alf}
\ea
where $a$ is a non-linearity parameter of the wave, defined here as a ratio of electric field
amplitude to background \Bf. Assuming linear shear,
$V_{z}= \eta y$, Eq. (\ref{Alf}) gives
\ba &&
E_{y} =B_0 \left ( - {\eta y \over c}  + a \sin k_{z} ( z- (V_A - \eta y)t) \right )
\nn  &&
B_{x}= B_0 \left ( 1 -  { a c \over  V_A - y\eta}  \sin k_{z} ( z- (V_A - \eta y)t) \right )
\ea 
where
$\eta = V'$ and we assumed that 
a test particle is located near the zero 
wave phase.

The equations of motion of a test particle are
\be
\partial_t \left( \gamma \partial_t {\bf r} \right) =
{Z e \over m c^2}  \left( \E +  { \partial_t {\bf r} \over \gamma c} \times \B \right)
\ee
Next, we separate particle motion into slow guiding center motion ${\bf R}(t)$ and
fast rotation $\xi (t)$
\be
{\bf r} = {\bf R}_0(t) + {\bf \xi}  (t)
\ee
with $|\xi|$ of the order of Larmor radius $ \sim r_L$. Assuming long wavelength
perturbations  $k_{z} \rightarrow 0$, weak perturbing wave $a \rightarrow 0$, 
and expanding in small gyro radii $r_L = \sqrt{\gamma_0 ^2 - 1} /\om_B$
(we neglect motion along \Bf, $p_{x}=0$),
we find equations of  motion
of the gyrations center, 
\ba &&
\partial_t^2 Y_0 = { Z \om_B \over \gamma_0} \partial_t Z _0
+ {Z \om_B \over \gamma_0}   { \eta Y_0  \over 2} 
- k_{z}  t ( Z \om_B V_A^2 /  \gamma_0 -\eta)
{ \delta E \over 2  B_0 V_A}  
\nn &&
\partial_t^2 Z_0 = - \left({Z \om_B \over \gamma_0} - \eta \right) \partial_t Y_0
+ k_{z} { \delta E \over 2  B_0 V_A}
\label{drift}
\ea
where capital letters denote coordinates of ${\bf R}_0$. Equations (\ref{drift})
can be integrated to give
 a drift in the $y$-direction with velocity
\be
{U_d \over c}  =  a  { c \over  V_A} 
{ k_{z} c \gamma_0 \over 2 Z \om_B} 
{ 1 - (V_A/c)^2 + \eta  \gamma_0  / (Z \om_B) \over 1-  \eta  \gamma_0  / (Z \om_B) }
\approx  a  { c (1 - (V_A/c)^2)  \over  V_A}
{ k_{z} c \gamma_0 \over 2 Z \om_B}
\label{Ud}
\ee
where we have reinstated the speed of light explicitly and the last
approximation assumes small $\eta \rightarrow 0$. 
In addition, there is  insignificant motion along $z$ which results in a $z$ displacement
of the order of $k_{z} \delta z \sim a  (c/V_A) \leq 1 $.

The origin of terms in Eq. (\ref{Ud}) can  be easily understood:
$ a  ( c /  V_A)$ is a relative amplitude
of \Bf perturbations due to presence of the wave, ${ k_{z} c \gamma_0 /2 \om_B} $ is a ratio
of particle Larmor radius to inhomogeneity scale and the last terms represent 
relativistic and shear corrections.
 Note, that as plasma becomes
strongly magnetized, $V_A \rightarrow c$, value of the
drift velocity decreases. This can be understood by noting that
for subluminal waves there is always a frame where electric field is zero
and \Bf is constant \citep{Clemmow}. What is important for drift
motion is the value of the \Bf in that frame. 
For shear \Alfven waves this frame is moving with the velocity 
$V_A$. For $V_A \rightarrow c$, the value of \Bf in this frame decreases
which leads to smaller transverse drift velocity. 

For order-of-magnitude estimates, we can assume
$V_A \sim c$.
[Note that for $V_A \ll c$, for a given potential
total energy flux increases by $(V_A/c)^2$, while for $V_A \rightarrow c$
drift velocity becomes small, so that the mechanism operates best at
$V_A \leq c$.]
Then
\be
{U_d \over c}  \sim a 
{ k_{z} c \gamma_0 \over 2 Z \om_B} 
\label{Ud1}
\ee

The energy of a particle changes both due to shear-induced electric field
as well as the electric field of the perturbing wave. 
We are interested in the former contribution; the  latter process -- 
gradient drift along the  
\Ef of the wave gives a well known surfing mechanism of particle
acceleration \citep{kats83}.
Energy gain in such a process is limited by electric 
field of the wave and spatial dimension of the wave.
Thus, if wave amplitude is of the order of unity, $\delta B / B \sim 1$
wave surfing will lead to  similar energy gains as
drift in a sheared flow. 
In astrophysical applications we expect $\delta B / B  \leq 1$, so that
 shear-induced potential 
is typically larger than the electric potential associated with perturbing
wave; see also \S \ref{surf} for more detailed discussion of effects of surfing.

Assuming that a particle keeps coherence with the wave (see \S \ref{surf}),
its energy in a shear-induced electric field 
 evolves according to
\be
\partial_t \gamma = { a^2 \over 4} { \gamma^2 c^3 t \over Z L_B^2 L_V\om_B}
\label{Edot}
\ee
where we introduced inhomogeneity scale of \Bf
$L_B  = 1/k_{z}$ and shear scale $L_V \sim c/V'=c/\eta$.
Note that variations of energy are independent of the sign
of \Bf gradient $L_B$, but depend on particle charge, \Bf and \Ef gradient.

Eq. (\ref{Edot})  implies very fast energy growth:
\be
\gamma = {\gamma_ 0 \over 1- (\gamma_0 /Z) (t/\tau_{0})^2}
\label{gammat}
\ee
where 
\be
\tau_{0} ={ 2 \sqrt{2} \over a } {L_B \over c} \sqrt{   L_V \over c/\om_B}
\ee
 Note that {\it the acceleration time scale is  {\it inversely}
proportional to the square root of initial particle energy:}
\be 
\tau_{acc} = {\tau_{0} \sqrt{Z} \over \sqrt{\gamma_0}}
\label{tauacc}
\ee
{\it 
Thus, more energetic particles with smaller charge are
accelerated on shorter time scales. }

Initially the energy grows quadratically with time.
At later times, the  Larmor radius of a particle starts to increase in response to  increasing energy.
This leads to faster drift and faster rate of  energy gain, which 
formally becomes super-exponential (finite time singularity).
Of course, in order to get infinite energy a particle should move infinitely far,
so that in reality  maximum energy gain will be limited by the size of the system and
it's total available potential.
In addition, 
before the finite  time singularity is reached, drift approximation will break down.
This will occur when the particle Larmor radius becomes of the order of inhomogeneity scales $L_B,\, L_E$.
This happens only a short time before finite time singularity, at times
\be
t= \tau_{acc} \sqrt{ 1 - {\gamma_0 c/\om_B \over Z (L_B, \, L_V)}} \sim \tau_{acc}
\ee

\subsubsection{Nonlinear waves: effects of wave surfing on energy gain}
\label{surf}

We have shown that if a particle can keep in phase
with the wave in a combined electromagnetic fields of sheared flow and inertial \Alfven wave,
  its energy increases at super-exponential rate. 
In case of a linear wave this assumption is not completely satisfied: 
 the acceleration time scale (\ref{tauacc}) 
is longer than the wave period by a factor 
$ \sqrt{L_V/ r_L}$.
 This difficulty may be overcome if perturbing wave is  non-linear, with finite
relative amplitude  $a=\delta E /B_0$.
In this case the particle will also experience  a drift in crossed
$\delta \E $ and $\B_0$ fields. 
If a particle is located at such phase of the wave so that this drift velocity
is along the direction of propagation, it may remain in resonance for a long
time. For substantially strong waves the longitudinal drift velocity 
(along the direction of wave propagation)
$u_{z} \sim \delta E /B_0$ may become of the order of the phase
velocity of the waves. Such particles are  effectively trapped by 
the wave. In surfing acceleration schemes this leads to particle
energy gain as it drifts along the \Ef of the wave. In our case
trapping leads to nearly constant  velocity drift along the
shear-induced \Ef.

\begin{figure}[ht]
\includegraphics[width=0.45\linewidth]{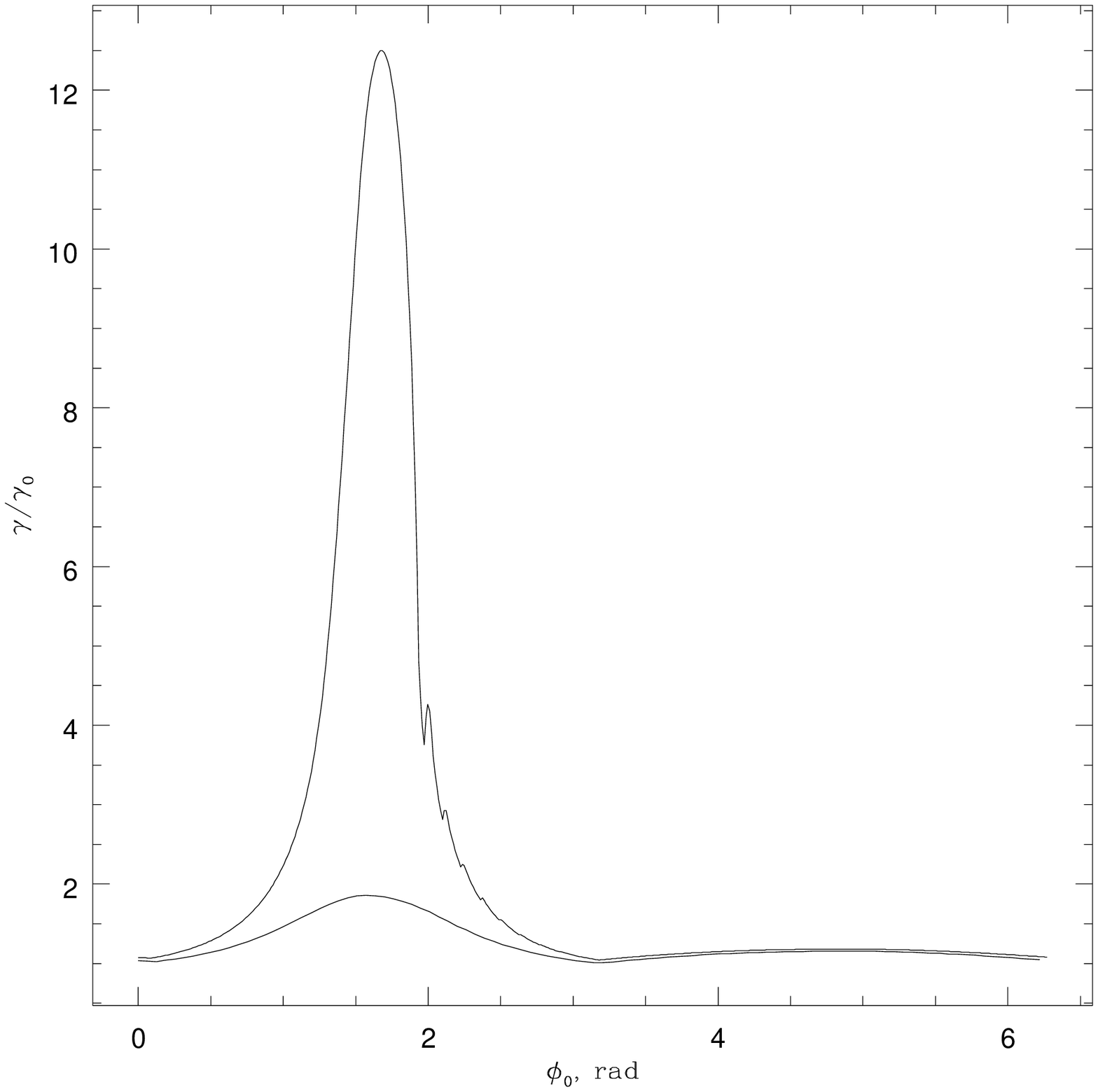}
\includegraphics[width=0.45\linewidth]{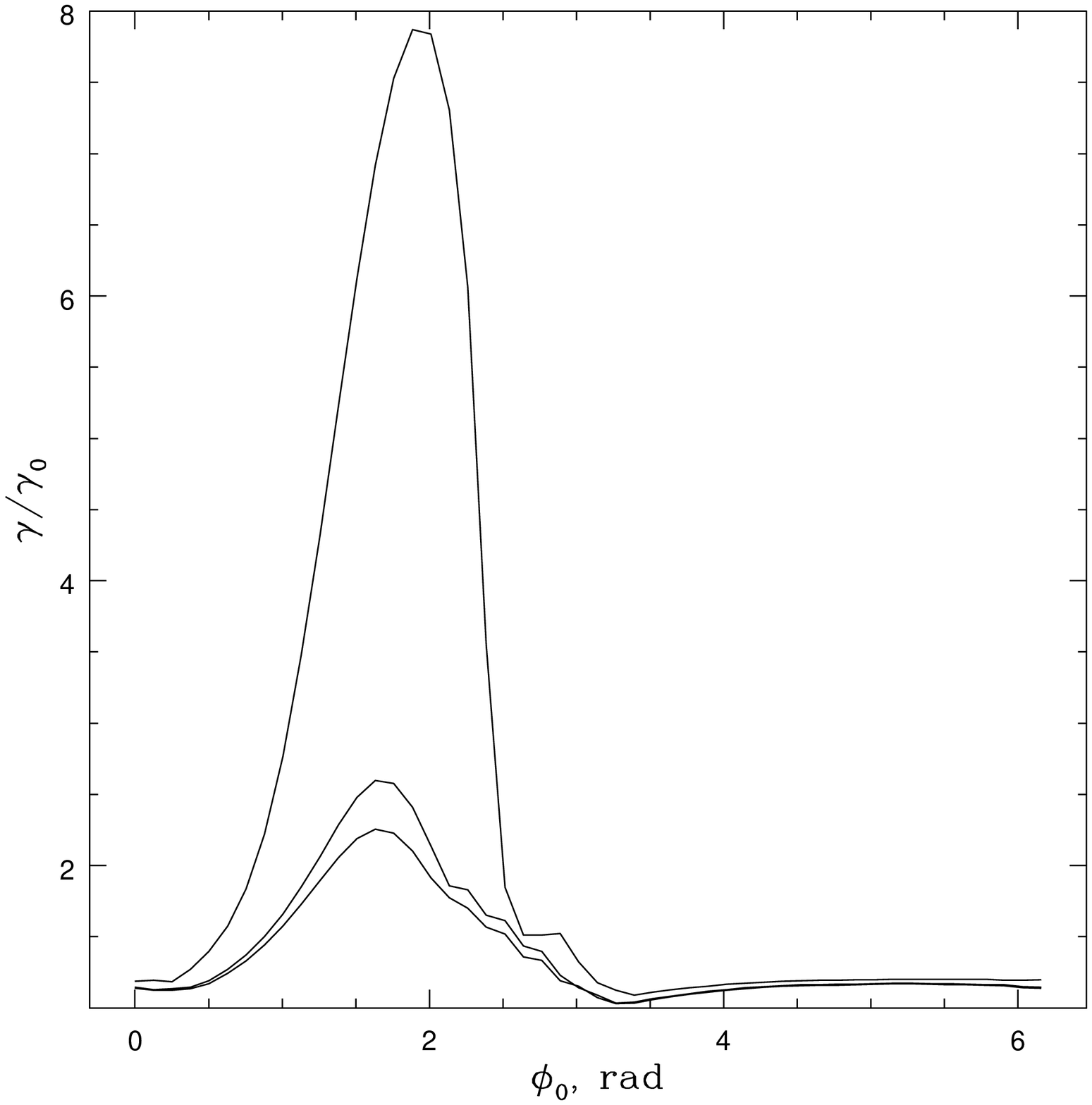}
\caption{ (a) Maximal  energy gained by a particle in a non-linear
inertial \Alfven wave propagating in a sheared flow as a function of initial phase
of a test particle
 for the following parameters: $V_A = 0.5 c$, $a=0.25$, $k_z=0.025$, $y_s=-5$
 (both in units $c/\omega_B$),
initial energy $\gamma_0 = \sqrt{2}$. Upper curve corresponds
to the full case shear plus a wave, lower curve is the case without
shear.  (b) Dependence of maximal energy gain on shear scale,  $y_s=-3$, $y_s=-10$,  $y_s=-15$
(top to bottom). In smaller shear (larger negative  $y_s$) acceleration is less efficient.
}
\label{gammamax}
\end{figure}

To test this possibility we consider numerically particles motion
in a sheared flow with finite amplitude inertial \Alfven wave. 
As a test case, we consider planar problem, similar
to the one described in Fig. \ref{shearCR}. For background velocity profile  we chose
$v_{z} =V_A \tanh ( y/y_s)$, where $y_s$ is a parameter determining the
strength of the shear. On top of the shear we impose an 
inertial \Alfven wave, with fluctuating fields given
in the WKB approximation (appendix \ref{waves})  by Eqns (\ref{Alf}).
We then seed the wave with test particle at  random  phases in the wave  
 and follow their trajectory.  
Results of simulations are presented in Fig. \ref{gammamax} 
where we plot the 
maximal  energy gained by a particle in a non-linear
inertial \Alfven wave propagating in a sheared flow as a function on initial phase
 for the following parameters: $B=B_{x}$, $V_A = 0.5 c$, $a=0.25$, $k_{z}=0.025$, $y_s=-5$
(negative $y_s$ is needed to have a negative  shear; for positive $y_s$ there
is no acceleration),
initial energy $\gamma_0 = \sqrt{2}$, initial momentum is directed along
$y$ axis. Upper curve corresponds
to the case of shear shear flow and  an inertial \Alfven  wave, 
lower curve is the case without
shear. Fig. \ref{gammamax} shows that 
particle indeed can gain large energies in a  sheared flow:
{\it most of the energy gain is  due to motion in sheared \Ef and not
due to wave surfing:  the wave mostly provides a drift velocity along
the shear}.  Without  a wave, particle just executes gyrational 
motion near the point $y=0$.

Another related type of acceleration in strong waves should be mentioned. 
In a strong inertial \Alfven wave propagating across \Bf,
fluctuating \Bf  of the wave  $\delta B_{x}$ at some wave phases reduces total
\Bf, so that total electric field (which is a sum of wave
\Ef and drift-induced field) may become larger than  the 
total \Bf.  In this case, there exists a reference frame where the 
\Bf is zero and only electric field is present, so that
particles will experience strong acceleration.
This type of acceleration is similar to surfing acceleration in strong
electrostatic (upper hybrid) waves \cite{kats83}. 
Since both charges of particles are accelerated in opposite direction,
such mechanism cannot be responsible for UHECRs, see appendix \ref{double}.
For the parameters of our simulations condition $E < B$ is always satisfied.

\subsection{Stage 2: large  Larmor radius}
\label{rL}

When the Larmor radius becomes comparable to shear scale particle dynamics changes
 qualitatively.  As we show below 
\cite[see also][]{ganguli88}, in case of  narrow  shear layers,
when Larmor radius becomes comparable to shear scale,
particle  motion becomes unstable even for
homogeneous  flow, without perturbing inertial \Alfven wave.

The easiest way to consider motion of a test particle in such a case
is using Hamiltonian formalism. Let's assume that \Bf is in the $x$-direction 
and that there is a shear velocity
along $z$ with $v_{z}(y)=\beta(y) c $ (Fig. \ref{shearCR}) and transform to the frame where  $v(0)=0$.
Hamiltonian of a particle in a sheared flow,
$H =  \sqrt{(m c^2)^2+ c^2 ({\bf P}-e {\bf A}/c)^2} + e \Phi$, where
$ {\bf P}$  is generalized momentum,  ${\bf A}$ is vector potential and 
$\Phi$ if electric potential, can be written
as 
\be
H=
m c^2 \left( \sqrt{ 1+ { p_{y} ^2 \over m^2 c^2} +
{Z^2 \om_B^2 (Y_g -y)^2 \over c^2} }  +
 { e Z \Phi \over m c^2} \right)
\ee
where $p_{y}$ is momentum along $y$ and  $Y_g = p_{z} / m \om_B $ is $y$
coordinate of the guiding center.
For linear shear the Hamiltonian  is
\be
H=
m c^2 \left( \sqrt{ 1+ { p_{y} ^2 \over m^2 c^2} +
{Z^2 \om_B^2 (Y_g -y)^2 \over c^2} }  +
 {Z \om_B \eta y^2 \over 2 c^2} \right)
\label{H}
\ee
 with the corresponding  
equations of motion 
\ba &&
\partial_t y = {p_{y}/m \over \sqrt{1+  { p_{y} ^2 \over m^2 c^2} 
+ {Z^2 \om_B^2 (Y_g -y)^2 \over c^2} } }
\nn&&
\partial_t p_{y} = - { y \eta Z  \om_B  m} 
+ { (Y_g -y) Z^2 \om_B^2 m \over  \sqrt{1+  { p_{y} ^2 \over m^2 c^2}
+ {Z^2 \om_B^2 (Y_g -y)^2 \over c^2} }}
\label{g1}
\ea
Let us assume that center of gyration is at $Y_g=0$.
After dimensionalizing, $p_{y}/m c \rightarrow p_{y}$, $y \rightarrow c y/Z \om_B$, $t \rightarrow \om_B t$,
$\eta \rightarrow \eta \om_B$,
Eq. (\ref{g1}) becomes
\ba &&
\partial_t y = {p_{y} \over \sqrt{1+   p_{y} ^2 
+ y^2  } }
\nn &&
\partial_t p_{y} =  - {y \over  \sqrt{1+   p_{y} ^2 
+ y^2  } } - y \eta
\ea
From conservation of energy
$\gamma_0= \sqrt{ 1+p_{y}^2 +y^2} + \eta y^2 /2$, we find that the  turn-around point, where
$p_{y}=0$ is at 
$y_{max}= { \sqrt{ 2 } \sqrt{ 1+ \gamma_0 \eta + \sqrt{ 1+ 2 \gamma_0 \eta +\eta^2}} / \eta}$.
In addition, there exists a critical trajectory, so that as time goes to infinity a particle
moves with constant velocity along $z$ axis.
In this case $\partial_t p_{y}=0$ and we find $y_{max}= \sqrt{1/\eta^2 -1}$.
Equating the two expressions for $y_{max}$
we find critical values of shear
\be
\eta _{crit} = \sqrt{\gamma_0^2 -1} - \gamma_0 \sim - {1 \over 2 \gamma_0}
\, \mbox{ for $\gamma_0 \gg 1$}
\ee
 and
$y_{max,crit}= \sqrt{ 2 \beta_0 /(1- \beta_0)}$, where $\beta_0 = \sqrt{1-1/\gamma_0^2}$. 
Thus, for $\eta < \eta _{crit} <0$, orbits are unbound and energy growth exponentially.
In the limit $\gamma_0 \gg 1$ this occurs when the shear scale $L_V = c/V'$ is half of gyro-radius,
$L_V= c \gamma_0 /2 \om_B$.

\begin{figure}[ht]
\includegraphics[width=0.9\linewidth]{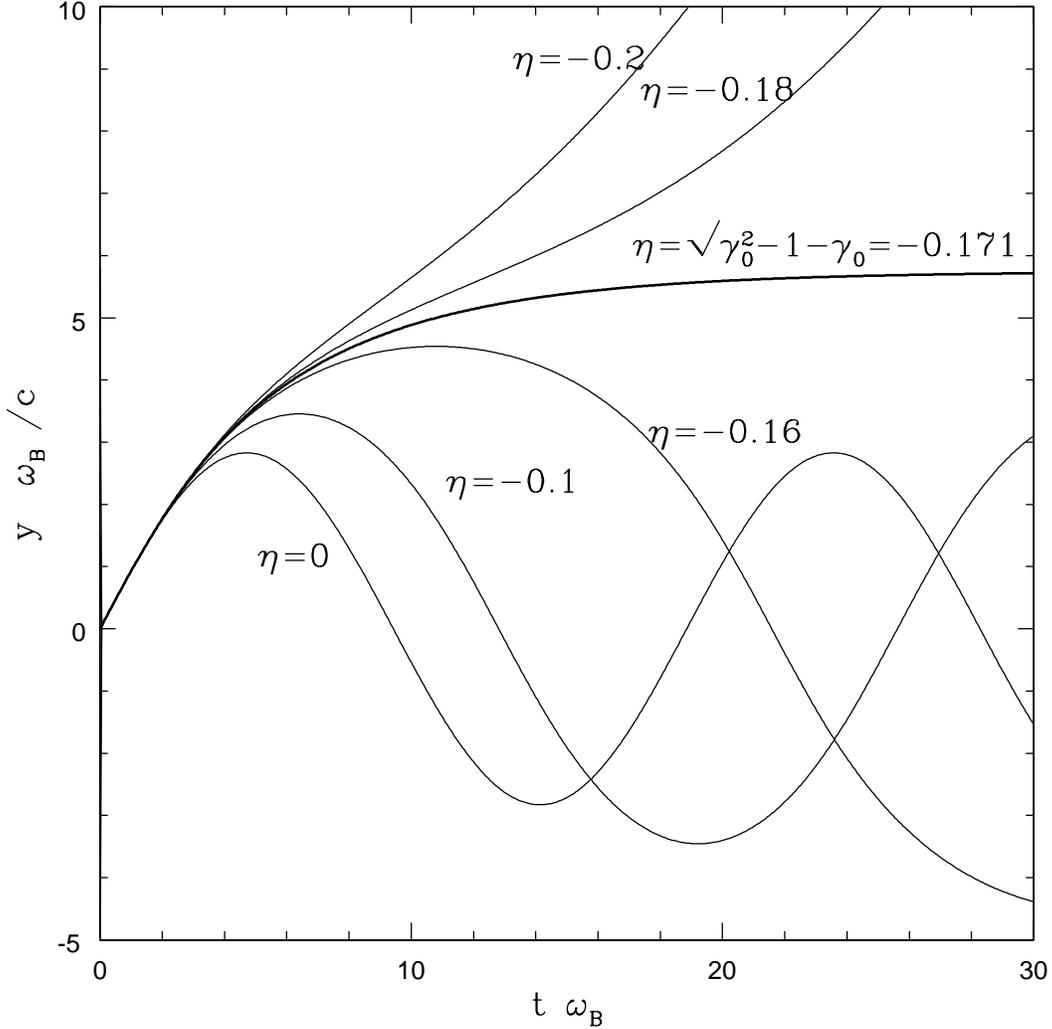}
\caption{
Particle trajectory $y(t)$ in strongly sheared layers with $v_z(y) = \eta y$ and for
 different values of $\eta$ measured in term of the Larmor frequency $\om_ B= e B_0 /  (m c)$. 
 Magnetic field  $B_x = B_0 \hat{x}$ is constant. 
 A particle initially at $y=0$,
with velocity along $y$ axes and Lorentz factor $\gamma_0=3$.
For $\eta < \eta_{crit} = \sqrt{\gamma_0^2-1} -\gamma_0 = -0.171 $ 
trajectories are unbound and particle  energy increases 
approximately exponentially, while 
for larger $\eta$   trajectories are bound.
Particle displacement is measured in term of non-relativistic Larmor radius
$c  / \om_ B$. For $\eta = \eta_{crit}$ particle reaches
$y_{crit} =  c /\om_B \sqrt{1/\eta_{crit}^2 -1}$ 
and moves along $z$ axis with constant
Lorentz factor $\gamma_{crit} = \gamma_0(1+ \eta_{crit} y_{crit}^2 \om_B/(2 c^2))$.
}
\label{etagrowthX}
\end{figure}

Particle trajectories
can be found in quadratures:
\be
t = \int 1/\sqrt{ 1 - {4  (1+y^2) \over (2 \gamma_0 - y^2 \eta)^2}} dy,
\ee
(see  Fig. (\ref{etagrowthX})).
In the  non-relativistic limit
equations of motion can be  integrated exactly,
\be
y= r_L \cos Z \om_B \sqrt{ 1 + \eta /Z \om_B} t, \,
z= - {r_L \over \sqrt{ 1 + \eta /Z \om_B}} \sin Z \om_B \sqrt{ 1 + \eta /\om_B} t
\ee
 This
 clearly shows that for strong negative  shear, $\eta < - \om_B$, particle trajectory is unstable and its energy growth exponentially.
For positive shear, $\eta > 0 $,  particle motion is stable.
Thus, when measured in term of particle gyration frequency $\om/\gamma_0$, 
critical shear reduces from $-1$ in the non-relativistic limit to $-1/2$ in the ultra-relativistic limit.

\begin{figure}[ht]
\includegraphics[width=0.9\linewidth]{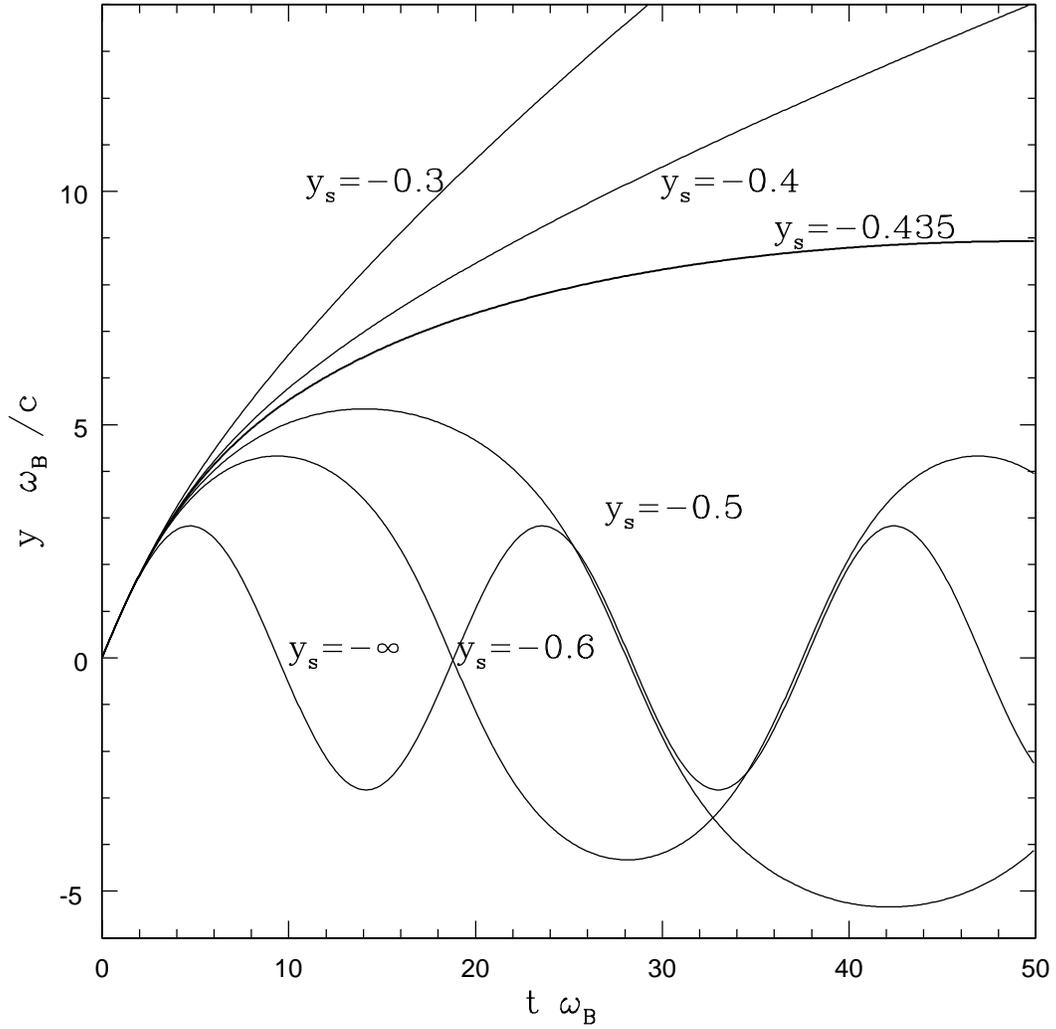}
\caption{  Same as Fig. \ref{etagrowthX} but for a velocity profile
$V_z = \tanh(y/y_s)$. For  $\gamma_0=3$, there is a critical values
$y_{s, crit}\sim -.435 $, so that for $ y_{s, crit}< y_s < 0$ particle trajectories
are unstable.}
\label{etatanh}
\end{figure}

Similar results may be obtained numerically for velocity profile
$V_{z} \propto {\rm tanh}(y/y_s)$. This case has an advantage over the linear profile since condition
$E<B$ is satisfied everywhere. For any given $\gamma_0$ there is a critical value
of $y_{s, crit} < 0$, so that for $ y_{s, crit} < y_s < 0$ particle trajectories are
unstable, see Fig. \ref{etatanh}.

Energy gain  occurs on time scale of $L_V / c \gamma_0$, which for strong shear
 is of the order of relativistic  Larmor frequency.
Thus, in this case {\it  acceleration rate reaches the absolute theoretical maximum.}

This acceleration mechanism can be contrasted with shock acceleration, in which case
acceleration stops when particle Larmor radius becomes of the order of system size. In contrast,
 inductive acceleration become more efficient when particle is unconfined.

\section{Spectrum of accelerated particles}

The acceleration mechanism proposed here requires a presence of pre-accelerated 
seed of particle with large Larmor radius.
Pre-acceleration mechanism may operate outside of the jet; radial drift then
may ''pull'' particles inside  a jet.
 Since \Bf at the periphery is relatively small
(\eg falling off as $1/r$), even relatively small energy particle
would have large Larmor circle and will experience fast radial drift. 
This can serve as an injection mechanism if at the base of the jet, near the
central galaxy, there is a population of pre-accelerated particles.

In a steady state, the total particle energy gain is determined by the
local electric potential, which, in turn, depends on \Bf and velocity
distribution. These  are  highly uncertain functions for the AGN jets. As an alternative
 model problem
we consider temporal evolution of distribution
of  pre-accelerated particles
 injected into jet at initial time and evolving  due to gradient drift
in sheared flow.
Assuming $\gamma \gg 1$, the evolution of the distribution function ($f=\frac{dn}{d\gamma}$)
in the rest frame is described by
\be
\partial_t f +  \partial_\gamma
\left(  \partial_t (\gamma ) f 
\right) = S
\ee
where $S$ is density of sources.  Assuming impulsive injection of seed population,
 integration along characteristics (\ref{gammat}) gives a solution
\be
f(\gamma,t) = f_0(\gamma_0(\gamma,t), t=t_0) \left({\gamma_0 \over \gamma} \right)^2
\ee
If the  initial injection spectrum is power-law, $f_0 \propto 1/\gamma_0^p$, then
\begin{eqnarray}
f(\gamma,t) &\propto&
{ 1\over \gamma ^p} \left(1+\gamma \left(  {t\over \tau_0}\right)
^2 \right)^{p-2} \\\nonumber
&\sim&  \gamma ^{-2} , \quad \mbox{for $\gamma t^2/\tau_0^2 \gg 1$}
\label{f}
\end{eqnarray}
Thus, for $p>2$  the spectrum flattens  with time.

The hardest spectrum that can be achieved has a power law index of 2.
This limiting case corresponds to unlimited acceleration in a  plane-parallel geometry,
which is realistically applicable to energies well
below the total available potential. At highest energies the final
spectrum will depend on the distribution of pre-accelerated particles with
respect to the electric potential and, in case of contribution from many
sources, on distribution of total potentials.

Relation between the observed spectrum and spectrum at the source is not
simple and depends on source distribution and (poorly constrained) extragalactic
\Bfs. Since high power AGNs are located at the edge of the GZK sphere,
 a flat spectrum at the source, with $p$ approaching 2,
 seems to be required in
order to account for AGASA observations of UHECRs above the ankle \cite{olinto00,olinto04}.
We note however that in the high energy region we
discuss here currently there seems to be a disagreement between the
AGASA ground array (\cite{takeda02}) and the HiRes fluorescence
detector (\cite{abbasi05}). Much larger experiments
such as Auger, Extreme Universe Space Observatory (EUSO), and 
 Orbiting Wide-angle Light-Collectors (OWL) will put much better constraints
on the injection spectrum. 

\section{Astrophysical viability}
\label{Astr}

One of the key observational test of UHECRs origin is the identification of the transition from the galactic to extragalactic components. The two competing models are the ''ankle scenario" \cite[see, \eg, ][for a general discussion of the transition]{demarco05}, when the transition is identified with the intersection of a steep galactic and flat extragalactic spectra, and the ''dip'' scenario, when  the transition is identified with the ''second knee"   \cite[see, \eg,][]{aloisio06}. 
Our model is consistent with the dip scenario as we discuss below.

{\bf Chemical composition}
In our model the EGCRs and the GCRs have 
different spatial and astrophysical
origin.  We assume however that the mechanism driving
the pre-accelerated component  at the acceleration cite of UHECRs 
is also at play in our galaxy.  From a pool of pre-accelerated particles, the ones with
 the highest rigidity are accelerated most 
efficiently.  Thus,  the population of pre-accelerated CRs is a mixture of heavy and light nuclei, the drift mechanism will preferably accelerate protons.
 Our model thus predicts an increase  
in proton over heavy elements as the energy of the UHECR
increases. Such trends have apparently been observed in the
composition of UHECRs with energies between $\sim 5\times 10^{17}$ eV
(''second knee'')
and $10^{19}$ eV and extending to energies
above $10^{19.3}$ eV (\cite{shinozaki,ave03}).
[See \cite{watson04}
  for a discussion of systematic differences between different measurements
 and how it limits our knowledge of the composition of CRs of the
highest energies.]

{\bf Injection spectrum}
The hardest  injection spectrum the model can account for is power law index of $2$. 
The dip models generally prefer softer spectra, with index $2.5 - 2.7$; such softer spectrum can be achieved both due to
spread out distribution of pre-accelerated particles  with respect to the total potential inside a given jet
and by a combination of contribution from many sources with different potential drops \cite{Kachelriess06,aloisio06}. For the case of injection spectrum with index $2$ the distribution of the
sources with respect to maximum acceleration energy $E_{max}$ required to explains the observational
data is $ dn/dE_{max} \propto ^E_{max}{-1.5}$  \cite{Kachelriess06}. The model also naturally provides a lower cut-off to the injection spectrum (see below).

{\bf Density of sources: isotropy and clustering of UHECRs}.
We associate acceleration cites of UHECRs with powerful FRII galaxies. Assuming proton composition,
to account  for energies above $10^{20}$ eV the source
 luminosity should be no less than $10^{46}$ ergs/s (see \S 2).  Assuming $\sigma \sim 1$, this translates into a limit 
on a total power of a jet.  The total jet power, as well as power in electromagnetic  component are  not straightforward to estimate.
Rawlings and Sauders \cite{Rawlings91} give estimates of jet power in FR II galaxies based on
assumption of equipartition between the \Bf and relativistic electrons and expansion dynamics of hot spots.  According to this {\it minimum} energy estimate, powers of FR II galaxies
reach $10^{45}-10^{47}$ erg/sec, while FR I are an order of magnitude less 
powerful.  This limits the possible acceleration sites to FR II radio galaxies  
which are typically far apart.  The power in strongly beamed sources, like  flat spectrum
radio quasar, are harder to estimate: radiative modeling 
\citep{ghisel03} gives  estimates comparable to FR II galaxies, up to
 $ 10^{49} $  erg/s.
In any case, requirements on luminosity (and assumption of high proton content
in UHECRs) {\it excludes} prominent low power nearby sources, like FR I radiogalaxy Cen A, nearest AGN M87 and starburst galaxy M82. We favor high power radio galaxies at intermediate distances 
as sources of UHECRs. 

The observed number density of highest power sources is a power law in energy, with sources
of $L \sim 10^{45} erg/s$ having density $\sim 10^{-6} Mpc^{-3}$. 
This is smaller by approximately an order of magnitude  than 
the density required to produce the small scale clustering,   $\sim 10^{-5} Mpc^{-3}$, \citep{blasi04}, but, first, the inferred density has a large uncertainty and, second,  the intrinsic electromagnetic luminosity is definitely higher than the observed  radiative one.

{\bf Shape of GZK cut-off}.
Small spacial density of sources also has  a clear prediction for the shape of the GZK cut-off
\cite{aloisio06}. Our model  predicts very steep GZK cutoff.

{\bf Total energy budget}.
 The model naturally avoids the energy budget problem posed by the steep required injection spectrum (in which case total energy density is determined by a somewhat arbitrary  lower energy cut-off). In the proposed mechanism, below some limiting energies particles just do not leave the jet. Though details depend on jet properties,   requiring that a cosmic ray crosses a 10 kpc jet over acceleration of jet length of 100 kpc,  one can estimate $E_{min} \sim 10^{16} $ eV.

{\bf High energy photon and neutrino traces}.
If high power, fairly distant AGNs are sources of UHECR, 
one may expect that as particle travel towards the observer 
they initiate electromagnetic cascades due to photo-pion 
production on CMB. Most of the energy of electromagnetic cascades ends up
in photons in the  GeV-TeV range.
If propagation is almost rectilinear, one may expect
an increase of high energy photon flux towards a source
at a level of $\sim 10^{-14}$ cm$^{-2}$ s$^{-1}$
\cite{fbdm04}. This flux can be (barely) 
detected by Cherenkov experiments like HESS.
Deflection of UHECR in localized
regions of high \Bf, associated, for example, with supergalactic plane 
may easily make this flux to be distributed over  large area of the sky.
To produce considerable deflection a magnetized sheath of thickness  of 1 Mpc
with \Bf of the order of $10^{-6}$ is required \cite{sigl}.
This may also explain non-detection of UHECR towards a particular AGN.
Finally, significant neutrino fluxes can be generated if protons
are accelerated up to $10^{21}$ eV energies \cite{wax09}.
We note however that the induced $10^{18}$ eV neutrinos are below the currently
advertised threshold for EUSO, OWL and ANITA, and most of the potential events will
go undetected  \cite{halzen02}.

\section{Discussion}
\label{discuss}

In this work we have considered a new mechanism of  acceleration of  UHECRs to
energies above the ankle
by inductive electric fields in sheared flows.  The  main points are: 

(i) {\it In a sheared flow protons are at a maximum
electric potential energy if  ${\bf B} \cdot \nabla \times {\bf v} < 0$,}
so that  kinetic  drift motion along the shear
will lead to fast energy gains.
If a jet is generated by rotating 
system of black hole-accretion disk with the velocity
falling off with cylindrical radius, then
the quantity  ${\bf B} \cdot \nabla \times {\bf v} $ will
be determined by the product $\B \cdot \Omega$ at the source, 
where $\Omega$  is the angular velocity of the black hole-accretion disk system
driving the jet.

Radial drift  motion may occur if a jet is inhomogeneous
along its axis, so gradient drifts will lead to energy gains.
Alternatively, one may imagine that the toroidal \Bf is perturbed azimuthally and is sheared by
differential velocity flow. This will generate non-axisymmetric  poloidal \Bf,
which would result in radial drift. Finally, particle diffusion, resonant or due to 
field line wondering, may lead to radial particle motion. In any case, a particle must cross the \Bf lines in order to  gain energy.


(ii)  Since drift velocity is an increasing function of energy, 
{\it acceleration rate is also an increasing function
of energy}. Thus, from a pre-accelerated population this mechanism
will pick up the highest energy particles and boost their energy further.
In addition, since energy gain is independent of the direction of the drift,
 pre-accelerated population may be located outside of the jet and be pulled-in 
by gradient drift. Pre-acceleration may be achieved, for example, by non-linear shock
acceleration in supernova and/or cluster shocks \citep{bell01}. 

(iii) After a particle has crossed a substantial fraction of the total
potential available in the system, its Larmor radius becomes comparable
to the flow scale. In this case exponential  acceleration 
may proceed without gradient drift.
 As a result, the {\it  acceleration rate reaches the 
theoretical maximum (acceleration on time scale of one gyration)}
and finally a particle can leave the system.

(iv) Since acceleration is fastest for highest rigidity particles,
from a population of a pre-accelerated particles with mixed composition,
{\it  particles with smallest charge -- i.e. protons -- are accelerated most efficiently}.
We thus predict a dominance of protons over heavy nuclei for energies above the ankle.

(v) Drift acceleration can also produce hard spectra, which are required 
if sources are located at the edge of the GZK sphere.
We have also calculated the spectrum of UHECRs assuming impulsive injection
of pre-accelerated particles with  a steep power law spectrum. {\it
Due to inductive acceleration the energy spectrum flattens and reaches 
asymptotically a power law index of 2. }

The best astrophysical  location for operation of the proposed mechanism is 
{\it cylindrically collimated}, high power AGN jets. The proposed mechanism cannot work
in spherically (or conically) expanding outflows since in this case 
 a particle  experiences polarization drift, which is a {\it first order} in Larmor radius,
 due to the fact  that in the flow frame magnetic field decreases with time. 
For a constant velocity flow this drift is always against the electric field (for a positively
charged particle) and lead to the decrease of energy on time
scale  $R/(c \Gamma)$, where $R$ is  a distance from the central source
and $\Gamma$ is the Lorentz factor of the flow. 
[A flow must acceleration faster than $\Gamma \sim r$ in order to beat this polarization drift.]
Comparing with acceleration time scale (\ref{tauacc}), and assuming $L_E \sim L_B
\sim R/\Gamma$, we conclude that the expansion time scale 
is shorter than the  acceleration time scale by a  factor
$\sim \sqrt{ R c / \gamma \Gamma \om_B} \gg 1$.
Thus, in spherically expanding flows adiabatic losses always dominate over
regular energy gain due to drift motion: the proposed mechanism 
would fail then.  On theoretical grounds,
AGN jets (or at least their cores) may indeed be asymptotically 
cylindrically collimated  \cite{HeyN03}. 
Observations of large scale jets, \eg Pictor A, do show jets that seem to be cylindrically
collimated on scales of tens of kiloparsec.

In cylindrically collimated parts of the jet acceleration can happen from sub-parsec to hundreds of
kiloparsec scales: as long as the motion of the jet is relativistic the total electric potential
remains approximately constant. Thus, UHECRs need not be accelerated close
to the central black hole where radiative losses are important \cite[\cf][]{levinson00}.
After a jet has propagated  parsecs from the central source
 radiative losses become negligible, see Eq. \ref{B}.

Another  constraint on the mechanism comes from the requirement that in order
 to produce radial kinetic drift the 
jet \Bf should be inhomogeneous along the axis. Though shocks provide
 possible inhomogeneity of \Bf ($\delta$-function inhomogeneity on the shock
front), we disfavor shock since particles are advected downstream and cannot
drift large distances along shock surface.
In a gradual inhomogeneity a particle drifts orthogonally to the 
field gradients and thus generally will remain in the region of 
inhomogeneous fields.
[One may imagine a hybrid model, where scattering  brings a particle
back to shock front, similar to Eichler's model of shock acceleration
near Earth bow shock, \cite{eichler81}].
Extragalactic jets are expected to have axial inhomogeneities,  both due to non-stationary 
conditions at the source and 
due to propagation of compression and rarefaction waves generated
at the jet boundary via interaction with surrounding plasma.

\section{Predictions}
\label{predict}

Our model has a number of 
clear predictions, some of which are related to astrophysical  association of acceleration sites of UHECRs with AGN jets \cite{olinto00} and some are specific to the model:
  (i)  one needs a relatively powerful AGN,
with luminosity $\geq 10^{46}$ erg/sec. This limits possible sources
to high power sources like FR II radiogalaxies,  radio loud quasars and high power BL Lacs
(flat spectrum radio quasars).
  Powerful AGNs are relatively rare and far
apart, so that a  steep GZK cut-off  corresponding to large source separation should be seen. 
(ii)   UHECRs come from sources with low spacial density. This may be reflected in the
distribution of arrival directions. 
(iii) Extragalactic  UHECRs   should  be  dominated by  protons.
(iv) Depending on ''extra-galactic seeing conditions'' arrival directions of  UHECRs may point to
their sources, though complicated \Bf structure may erase this correlation.
In addition,
the fact that only flows with negative shear can accelerate protons
implies that only approximately half of such AGNs can be sources of UHECRs (this assumes that the AGN central engine - black hole or an accretion disk - is  dominated by  large scale, dipolar-like \Bf).

We would like to thank  G. Sigl  for discussions and comments on the manuscript.
We also thank R. Blandford, C. Dermer, J. Heyl, M. Medvedev,
M. Ostrowski and A. Zhitnitsky. 
This research is supported by a grant from the
Natural Science and Engineering Council of Canada.



\appendix

\section{Impossibility of acceleration of UHECRs in DC electric fields}
\label{double}

Acceleration by  DC electric fields, when  there is component of \Ef along \Bf
or when $E>B$, is fast and very 
efficint way of gaining energy. On the other hand,  in this appendix we argue
that  {\it UHECRs cannot be accelerated by DC electric fields}.
There are two reasons.
First, if we have a  gap (region where there is an uncompensated \Ef
parallel to \Bf), any lepton will undergo acceleration and after passing a potential
drop of the order of $10^6-10^8$ eV will produce a pair, \eg through inverse Compton
effect or curvature emission.
The secondary pairs will similarly be accelerated, producing an electromagnetic cascade
that will short out the initial DC  \Ef. 
Secondly, even if one manages to create an extremely  photon-clean surrounding with straight
\Bfs,  the \Ef will accelerate both signs of  charges, producing field-aligned current.
The magnetic field, associated with these currents would  have prohibitively  large energy 
as we show below.

Consider a one dimensional double layer with potential difference $\Phi_{tot}$ that can freely emit particles from each end. To account for possible effects of rotation and associated 
charge density  we introduce  Goldreich-Julian density $n_{GJ}$ in the potential equation
The governing equations are
\ba &&
\partial_z^2  \phi =  4 \pi (n_i e_i -n_e e+ n_{GJ} e)
\nn &&
j_e = e n_e v_e =const \nn &&
j_i = e_i n_i v_i  =const \nn &&
\left( {1 \over \sqrt{1-v_e^2/c^2}} -1 \right) m_e c^2 =  e \phi
\nn &&
\left( {1 \over \sqrt{1-v_i^2/c^2}} -1 \right) m_ic^2 =  e_i( \Phi_{tot} -  \phi)
\ea
where $z$ is the coordinate along the electric field,
$n$ and $e$ are density and charges, $j$ is the current density, $v$ is velocity of two species.  Thus
\ba && 
v_e = { \sqrt{ \phi} \sqrt{ 2 m_e c^2 /e + \phi} 
\over m_e c^2 /e + \phi } c
\nn &&
v_i = { \sqrt{ \Phi_{tot} -  \phi}  \sqrt{ 2 m_ic^2 /e_i +\Phi_{tot} -  \phi} 
\over  m_ic^2 /e_i +\Phi_{tot} -  \phi}c
\ea
Putting into  potential equation, we find
\be
\phi'' = {4 \pi \over c} 
\left( n_{GJ} e - {  j_e ( e \phi + m_e c^2 )\over \sqrt{ e \phi  ( e \phi + 2 m_e c^2)} }
+ {j_i (m_i c^2 + e_i ( \Phi_{tot}- \phi)) \over 
 \sqrt{ e_i ( \Phi_{tot}- \phi) ( 2 m_i c^2 + e_i ( \Phi_{tot}- \phi) ) }   } \right)
\label{pp}
\ee
which can be integrated once. The solution of (\ref{pp}) should satisfy 
 the boundary conditions: at $\phi(z=0)=0, \,\phi' (z=0)=0$,  $\phi(z=H)=\Phi_{tot},\, \phi'(z=H)=0$  
(there is no electric field
on each end). 
Thus the problem is an eigenvalue problem on currents. We find
\be
{j_i e \over j_e e_i} = \sqrt{ {  2 m_e c^2 /e + \Phi_{tot} \over 2 m_ic^2 /e_i +\Phi_{tot}}}
+ {n_{GJ} e_i c \over j_i \sqrt{ 2 m_e c^2 / ( e \Phi) +1 }}
\ee
which generalizes  the 
relativistic Child's law  \cite{raadu}, see also \cite{arons79}.
Equation for $\phi'$ then becomes
\ba &&
{ \phi^{\prime2} \over 2} = 
{ 4 \pi j_e  \over c } 
\left( \sqrt{ 2  m_e c^2 /e + \Phi_{tot}} 
\left( \sqrt{ \Phi_{tot}} - {  \sqrt{ \Phi_{tot} -  \phi} \sqrt{ 2 m_ic^2 /e_i +\Phi_{tot} -  \phi}
\over \sqrt{ 2 m_ic^2 /e_i +\Phi_{tot}}} 
\right) \right.
\nn &&
\left.
- \sqrt{ \phi}  \sqrt{ 2 m_e c^2 /e + \phi} 
 -
{n_{GJ} e c \over j_e} 
\left( \Phi_{tot}-\phi - \sqrt{ \Phi_{tot} ( \Phi_{tot}-\phi)(2 m_i c^2/e_i + \Phi_{tot}-\phi) \over
{2 m_i c^2/e_i+ \Phi_{tot}} }  \right)
\right)
\ea 
dimensionalizing
$m_e = \mu m_i$,  $\phi \rightarrow \phi \Phi_{tot}$,
$\Phi_{tot} = \gamma_0 m_i c^2 /( Z e)$ ($\gamma_0$ has a meaning of a maximal Lorentz factor
that ions can reach), $n_{GJ}  \rightarrow \beta_{GJ} j_i/ (e  c)$  and 
$z \rightarrow \gamma_0^{3/4}  \sqrt{ m_i c^3 \over 4 \pi e Z j_i} z$,
equation for $\phi'$ becomes
\ba && 
{ \phi^{\prime,2} \over 2} =
-\sqrt{\gamma_0+2 } + \sqrt{(1-\phi)(2+ \gamma_0(1-\phi))} + \sqrt{ (\gamma_0+2) \phi (2 Z \mu + \gamma_0 \phi) \over \gamma_0+2 Z \mu } - 
\nn &&
\beta_{GJ} { \sqrt{\gamma_0} \phi  } 
\left(1-\sqrt{ 2 Z \mu + \gamma_0 \phi \over \phi ( \gamma_0+  2 Z \mu)} \right)
\ea
In the limit of highly relativistic ions, $\gamma_0 \gg 1$,
this becomes
\be
{ \phi^{\prime,2} \over 2} =
{ \phi + (Z +\beta_{GJ}) \mu (1- \phi) \over \sqrt{\gamma_0} } \approx { \phi \over \sqrt{\gamma_0} }
\ee
This determine the length of the double layer in dimensionless units.
\be
z = \sqrt{2} \gamma_0^{1/4}
\ee
Returning to dimensional notations we find that the maximum potential
 a double layer of size $R$, carrying a current $I$ (so 
that $j_i \sim I /2 R^2$, assuming that the width of the current 
layer is of the order of its length) may  have is
\be
\Phi_{tot}= \sqrt{ 2  \pi} \sqrt{ I  m_i c \over Z e } 
\ee
The maximum energy is then
\be
{\cal E}_{max} \sim   \sqrt{I m_i c Z e  }
\ee
Since $I < c \sqrt{ E_{tot}/R}$, where $E_{tot}$ is total energy in a system,
\be
{\cal E}_{max} \leq  \sqrt{  m_i c^2 Z e }  \left({ E_{tot} \over R } \right)^{1/4} 
\leq 10^{15} {\rm eV}   \sqrt{ Z}  \left({ E_{tot} \over 10^{60} {\rm erg} } \right) ^{1/4}
  \left({ R \over 10  {\rm kpc} } \right) ^{-1/4}
  \label{Em}
\ee
Thus, the maximum energy achievable by a particle scales only as $ E^{1/4}$ 
(c.f. with maximum energy achievable in inductive field, $\propto L^{1/2}$, Eq. (\ref{phi})). 
Even the largest energy reservoirs, lobes of radio galaxies, cannot provide
large enough potential for DC acceleration of UHECRs.
For magnetars with $B\sim 10^{15}$ G and $R\sim 10^6$ cm estimate
(\ref{Em}) gives ${\cal E}_{max} \sim 1.6 \times 10^{16}$ eV,
while for supermassive  blackholes with $B\sim 10^4$ G and $R\sim 10^{13} $ cm, 
${\cal E}_{max} \sim 1.6 \times 10^{14}$ eV.
This excludes the possibility of UHECR acceleration by DC \Efs
in astrophysical environment.

There is only one way to circumvent this argument: acceleration region must be really 
devoid of any plasma \citep[as, for example, suggested by][in  the immediate surrounding
of dead quasars]{boldt99,levinson00}. 

\section{Can GRB shocks accelerate UHECR?}
\label{grbs}

Simple estimates (\ref{phi}) shows that only very luminous  sources have enough potential
to acceleration  UHECR. GRBs, with peak luminosity of the order of $10^{51}$ erg/s
and  total potential $\Phi_{GRB} \sim 4 \times 10^{22}$ V,
 are then  prime suspects \cite[eg][]{wax04}. Still, we disfavor this possibility
 for the  following reasons. Internal shocks, where UHECRs are supposed to be accelerated,
 are short transient events occurring at distances $\sim 10^{12}-10^{13}$ cm.
 In order to overcome radiative losses, from Eq. (\ref{B}) it follows that
 Lorentz factor should be $\Gamma \geq 300$.
 The first problem is that for spherically expanding ultra-relativistic  flows the 
 available potential is smaller than (\ref{phi}) by a factor $1/(\Gamma \Delta \theta ) \ll 1$,
 since particle can cross sideways only a distance $\sim R/\Gamma$.
 To see this, consider a conically expanding flow with total luminosity $L$
 and opening solid angle  $\Delta \Omega = \pi \Delta \theta^2$.
 [All the relations below apply, roughly, both to large scale and small scale \Bfs.] If $b$ is the \Bf
 in the flow rest-frame, using  (\ref{L}) we can write the rest-frame \Bf as
\be
b \sim \sqrt{  {\sigma \over 1+ \sigma} }
\sqrt{ L \over  c } {1 \over  \Gamma  \Delta  \theta R}
\ee
 The available potential in the rest frame is then
\be
 \Phi_{rest} \sim  b {R \over \Gamma}
 \sim \sqrt{   \sigma \over  (1+ \sigma)}
 { 1 \over \Gamma^2 \Delta \theta } \sqrt{  L \over c}
\label{Phirest}
\ee
 while in the laboratory frame it is
\be
\Phi_{lab} \sim \Gamma  \Phi_{rest} \sim \Gamma b {R \over \Gamma}
\leq \sqrt{  \sigma \over  (1+ \sigma)}
{ 1 \over \Gamma \Delta \theta } \sqrt{  L \over c}  =
3.6 \times 10^{21}\, {\rm V} \sqrt{  \sigma \over  1+ \sigma} 
\left( {\Gamma \Delta \theta \over 30} \right)^{-1} 
\left( { L \over 10^{51} {\rm erg s}^{-1}} \right)^{1/2}
\label{GRB}
\ee
(since both \Ef and \Bfs are larger in laboratory frame by  a factor $\Gamma$).
At the time when internal shocks are supposed to occur,  $\Gamma \Delta \theta \gg 1$,
so that the available potential (\ref{GRB}) is smaller than the total
potential by a factor $\Gamma \Delta \theta \sim 30$
(for $\Gamma =300$ and $\Delta \theta \sim 0.1$).
Potential (\ref{GRB}) is still large enough, so that sub-equipartition fields,
$\sigma \geq 0.01$ are sufficient.

[In passing we note that the restriction discussed above applies
 only to UHECR acceleration at GRB internal shocks
 occurring at small distances. Since the total potential remains approximately constant
 as long as the expansion is relativistic, $\Gamma \geq 1$, but the available potential
 increases for decreasing $\Gamma$ until $\Gamma \sim 1/\Delta \theta$ and remains constant
 after that, it is feasible that UHECRs are accelerated at later
 stages, when the flow becomes trans-relativistic. This cannot, though,  be done neither at external
 shock since magnetization there is low, nor at internal shocks, since the flow becomes
  trans-relativistic long time after internal shocks have been dissipated.]

Secondly, and most importantly, as long as particle has not crossed
{\it all}  the  available potential (\ref{GRB}), its Larmor
radius is smaller than the size of the flow in its rest frame $r_L < R/\Gamma$.
After crossing fraction $\xi$ of the  total available  potential in rest frame (\ref{Phirest}) the
 maximum Larmor radius  becomes
    \be
    r_L = \xi {  \Phi_{rest} \over  b}  = \xi {R \over \Gamma}\ ,
    \label{zeta}
    \ee
In a framework of shock acceleration, if Larmor radius is smaller than the size of the 
system, $\xi < 1$, a particle is advected with the flow and is then subject to
 strong adiabatic losses, losing all the energy that it gained
 from acceleration process. Thus, in order to avoid adiabatic losses a particle
 {\it must}  cross all the  available potential so that its Larmor
radius would become of the order of the flow scale and adiabatic approximation would break down.
 Condition (\ref{zeta}) is equivalent to the requirement  that
 acceleration time scale for highest energy UHECRs is of the order of flow expansion.
 Thus, the paradigm of shock acceleration of UHECRs in GRBs {\it requires }
 that  particles are accelerated on time scale of
 inverse relativistic gyro-frequency, nothing less will do.
We consider this condition, though formally within the limits of shock acceleration, as
a very restrictive and thus disfavor UHECRs at internal shocks in GRBs.

\section{Electromagnetic waves in transversely sheared plasma}
\label{waves}

In this Appendix we consider properties of electromagnetic  waves in transversely sheared cold plasma. 
The results presented here  will server as a basis for background
electromagnetic field that lead to particle  acceleration.
One of the key differences between our approach  
 and previous works 
\cite{ganguli88} is that {\it 
waves in transversely sheared plasma cannot be considered self-consistently in the framework
of non-relativistic MHD}, 
since, as we showed in Section \ref{motion}  
shear introduces charge density, which is neglected by
 non-relativistic MHD.
In addition, for perturbations propagating 
orthogonally to \Bf, {\it inertial \Alfven waves } play a crucial role, so that
induction current, which is  neglected 
in non-relativistic MHD, is important.

Consider cold plasma composed of two species  $\alpha=e,p$.
Maxwell equations, continuity and equations of motions then read
(in this Section we assume $c=1$)
\ba && 
\partial_t \B = - \curl \E
\nn &&
\curl \B = {4 \pi } \J + \partial_t  E
\nn &&
\div E = 4 \pi \sum n_\alpha  e_\alpha
\nn &&
\div B=0
\nn &&
\partial _t n_\alpha + \nabla ( \v  n_\alpha) =0 
\nn &&
 \J = \sum n_\alpha  e_\alpha \v_\alpha
\nn &&
\partial _t  \v_\alpha +  (\v_\alpha \cdot \nabla) \v_\alpha= 
{ e_\alpha \over m_\alpha} \left( E+ \left( { v _\alpha } \times \B \right) \right)
\ea
Below we assume that electrons (subscript 1) carry a charge $-e$ while 
positive chargers are protons with charge $e$.

Assume that the \Bf is along $x$-axis $B_{x}=B_0$ and that there is sheared flow
along $y$-direction with $V=V_{z}(y)$.
Then, there is electric field $E_{y}= - V_{z} (y) B_0$
and corresponding charge density
\be
\rho_0 = {1 \over 4 \pi} \div E= - {1 \over 4 \pi}  V' B_0
\ee

Consider small fluctuations on this background $\propto \exp\{i  ( k_{x} x + k_{z} z - \om t)\}$,
$\B = \B_0+ \delta b$, $\E = \E_0 +  \delta e$, $n_\alpha= n_0 +  n_\alpha$, ${\bf v} = 
{\bf V} + {\bf v}_\alpha$.
 Assuming,
for simplicity
  $k_{x}=0$, perturbations are governed by equations
\ba &&
\delta b_{y} = -{ k_{z} e_{z} \over \om} ,
\delta b_{z} ={i e_{x}' \over \om} ,
\delta b_{x}= - { ie_{z}' + k_{z} e_{y} \over \om} 
\nn &&
i \om e_{y} + 4 \pi n_0 e ( v_{1y} - v_{2y})=0
\nn &&
i \om e_{z} - \delta b_{x}' =  4 \pi n_0 e ( (n_1 - n_2 ) V + n_0 (v_{1z}-v_{2z}) )
\nn &&
\delta b_{z}' = i (k_{z} \delta b_{y} - \om e_{x})
\nn &&
\delta b_{y}' = - i k_{z} \delta b_{z}
\nn &&
e_{y}' + i k_{z} e_{z} =  4 \pi e( n_1 - n_2)
\nn &&
-\om n_\alpha + k_{z} n_0 v_{1x} - i n_0 v_{\alpha y}' +k_{z} n_1 V=0
\nn &&
\om_{B \alpha} ( e_{y} + \delta b_{x} V) +B_0  \left( i \om_D v_{\alpha y} + \om_{B,\alpha} v_{\alpha z} \right)=0
\nn &&
\om_{B \alpha} e_{z} + i  \left( \om_D v_{\alpha z} + v_{\alpha y} ( \om_{B \alpha} +V') \right)=0
\nn &&
e_{x} = \delta b_{y} V
\ea
where $\om_D = \om - k_{z} V$.
Eliminating $ \delta {\bf b}$ and $n_\alpha$:
\be
n_\alpha= { k_{z} v_{\alpha z} - i v_{\alpha y} ' \over \om_D}
\ee
we derive the velocity fluctuations as
\ba &&
v_{\alpha y} = { \om_{B \alpha} \left( \om  \om_{B \alpha} e_{z} - \om_D V e_{z}'- i \om_d^2 e_{y} \right) \over B_0 D_\alpha \om}
\nn &&
v_{\alpha z} = - { \om_{B \alpha} \left( i ( \om \om_d e_{z} - V e_{z}' ( \om_{B \alpha}+V') ) +
 \om_d e_{y}  ( \om_{B \alpha}+V') \right) \over B_0 D_\alpha \om}
\label{vel}
\ea
where $ D_\alpha = \om_D ^2 - \om_{B \alpha}^2  - \om_{B \alpha} V'$.
Equation for $e_{x}$ separates
\be
e_{x}'' = (\om^2 -k_{z}^2) e_{x},
\ee
while 
equation  $e_{y}$ and $e_{z}$  obey a system of equations
\ba &&
e_{y} = { i  \over \om^2 -k_{z}^2} \left(   4 \pi e n_0 \om ( v_{1y}-v_{2y})  +
 k_{z} e_{z}' \right)
\nn &&
ik_{z} e_{z} + e_{y}' = { 4 \pi e  n_0 \over \om_D}
\left(k_{z} (v_{1z}-v_{2z}) - i (v_{1y}' - v_{2y}')\right) 
\label{sys}
\ea
where velocities can be eliminated using (\ref{vel}).
In full generality the system  becomes
too complicated to be reproduced here. We consider two simple case:
propagation along $y$-axis and a low frequency limit
$\om \ll \om_{B\alpha}$.

\subsection{Propagation along $z$ axis}

For  wave with $e_{z}=0$, we can immediately 
find dispersion relation for the fast mode
(inertial \Alfven mode) polarized along $y$:
\be
\om^2 = k_{z}^2 + {4 \pi n_0 \om_D ^2 \over B_0}
\left( { \om_{B1} \over 
 \om_{B1}^2 + \om_{B1} V' - \om_D ^2}  - { \om_{B2}  \over
 \om_{B2}^2 + \om_{B1} V' - \om_D ^2} \right)
\ee

We are interested in low frequency waves $\om \ll \om_{B1}, \, \om_{B2} $.
Assuming that plasma is electron-ion, (first type of particles are electrons with $|\om_{B1}|
 \gg \om_{B2} $),
and introducing magnetization parameter
\be
\sigma = {\om_{B2}^2 \over \om_{p2}^2}
\ee
we find
\be
\om^2 = k_{z}^2 + { \om_D ^2 \over \sigma}
\label{D}
\ee
In the absence of shear this gives dispersion relation for inertial \Alfven waves
$
\om =  \sqrt{ \sigma /( 1+ \sigma)} k_{z} c =  V_A k_{z}
$
where $ V_A = \sqrt{ \sigma / 1+\sigma} c$ and we have reinstated explicitly the speed of light.
This relation is relativistically exact;
in the non-relativistic limit $\sigma \ll 1$, this gives
so called inertial \Alfven waves
with $V_A = \sqrt{\sigma} c= B /\sqrt{4 \pi m_p n_0}$.

In the non-relativistic limit $V_A, \, V \ll 1$
Drift dispersion relation (\ref{D})
gives 
\be
\om = k_{z} ( V_A \pm V)
\ee

\subsection{Oblique propagation normal to the \Bf: low-frequency limit}

Next we consider low frequency waves propagating obliquely to 
velocity shear.
In the  limit $\om \ll \om_{B2} \ll \om_{B1}$
 we find
\ba && 
v_{1y} - v_{2y} =   -
\left( {  \om_D ( k_{z} \om_D - (\om^2 - k_{z})^2 V) e_{z} ' - 
\om (\om^2 - k_{z}^2) e_{z} V' \over B_0 \om \om_{B2} ( \om^2 -  k_{z}^2 + \om_D ^2/\sigma) } \right)
\nn &&
v_{1z}- v_{2z} = { i \om_D e_{z} \over B_0 \om_{B2} }
e_{y} = {i \over \sigma}  { \om V' e_{z} + (k_{z} \sigma + \om_D)
 \over \om^2 -  k_{z}^2 + \om_D ^2/\sigma} 
\ea
which gives an equation for continuous spectrum  normal  modes
\ba &&
\left( { 1+\sigma -V^2 \over \sigma( \om^2 - k_{z}^2) + \om_D^2 } \right)
e_{z} '' 
+ 2  { (\sigma \om + \om_D) (k_{z} - \om V) V' \over  (  \sigma( \om^2 - k_{z}^2) + \om_D^2)^2 }
e_{z}'
\nn &&
+ 
\left( {1 +\sigma \over \sigma} + 2 { k_{z} (k_{z} - \om V) V^{\prime 2} \over  (  \sigma( \om^2 - k_{z}^2) + \om_D^2)^2 } \right) +  { (k_{z} - \om V) V'' \over \om_D  (  \sigma( \om^2 - k_{z}^2) + \om_D^2)}
\ea
Assuming weak non-relativistic  shear $V \ll 1$ (neglecting quadratic terms in V),
this reduces to
\be
 \partial _y \left( { 1+ \sigma \over \sigma( \om^2 - k_{z}^2) + \om_D^2 }  e_{z} ' \right) +
\left( {1 +\sigma \over \sigma} + {k_{z} V'' \over \om_D ( \sigma( \om^2 - k_{z}^2) + \om_D^2 )}  \right) e_{z}'' =0
\label{Rey}
\ee
This equation is  reminiscent of Rayleigh equation in compressible fluid \cite{LandauIV}:
second derivative of the velocity profile is divided by the shifted drift frequency.
This determines the usual  Rayleigh instability.
(For a flow to be  Rayleigh unstable, the necessary condition is presence of inflection point
where $V''=0$.).
[Also note, that for $k_{z}=0$ Eq. (\ref{Rey}) gives
$\om = V_A k_{y}$, so that shear does not affect wave propagation in this case.]

For linear profile $V''=0$ and in the non-relativistic limit $V \ll 1$,
Eq. (\ref{Rey}) reduces to
\be
\partial _y  { V_A^2 \over 
\om_D^2 - k_{z}^2 V_A^2}  e_{z} ' + e_{z} =0
\ee
Far from the resonance points 
equation of the form
\[
\partial _y \left( f(y) e_{z} ' \right) + e_{z} =0
\]
with slow varying function $f(y)$ can be solved by WKB method, which gives
\be
 e_{z} \propto { 1 \over \sqrt{f}} \exp\{\pm i \int dy/\sqrt{f(y)} \}
\ee

Properties of small perturbations on the background flow
are closely related  to the stability of the flow.  Ideal sheared flows
 may be subject to Kelvin-Helmholtz and Rayleigh instabilities
\cite{LandauIV}. In addition, in warm plasma new
types of instabilities appear, which are related to ion cyclotron motion
\cite{ganguli88}. Discussion of the stability properties of the
transversely sheared flows is beyond the scope of the paper: we assume
that the sheared  configuration under consideration is stable. 
This can be shortly justified by noting that
 Rayleigh-type instability depends on the presence of inflection point in the
 flow and can be eliminated by choosing, \eg, linear velocity profile,
 while kinetic instabilities are negligible if the background plasma
 is cold.

\newpage

\begin{figure}[ht]
\includegraphics[width=0.9\linewidth]{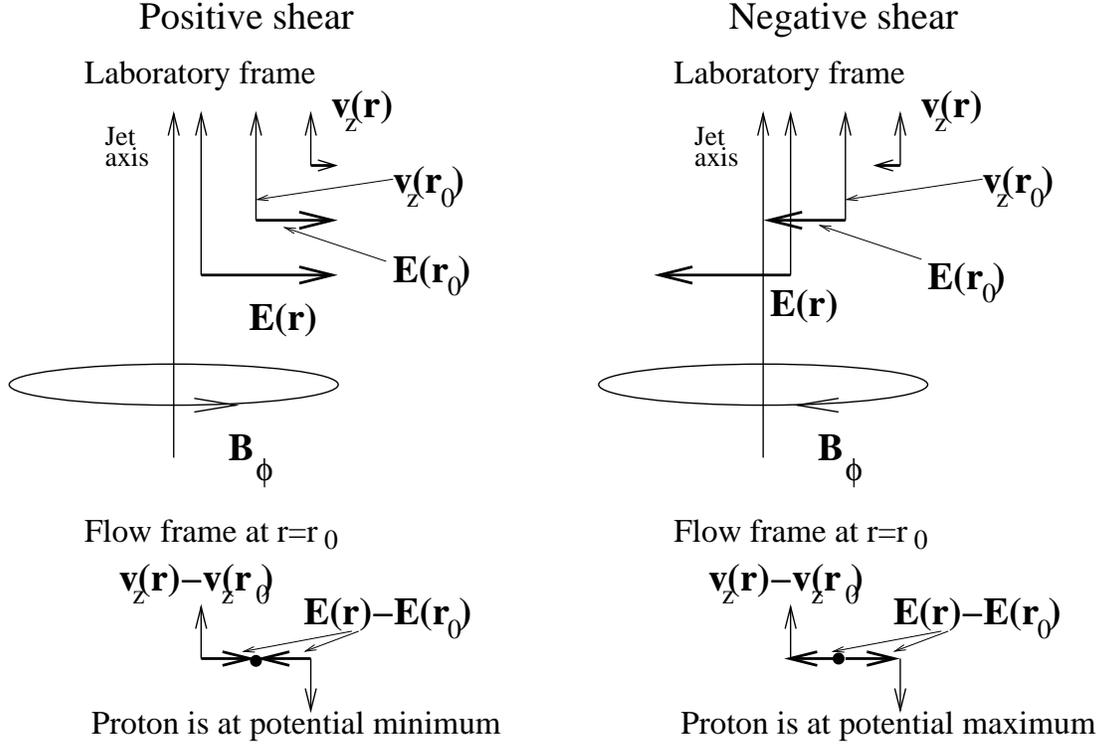} 
\caption{Potential energy of a charge in a flow for positive and negative shear.
Cylindrically collimated flow carries toroidal \Bf $B_\phi$   and is sheared, $v_z(r)$, so that
regions closer to the axis move with larger velocity. There is an inductive
electric field, directed either away from the axis (for positive shear) and towards the
axis (negative shear). In a rest frame moving with the flow at some $r=r(r_0)$ the electric field
is either towards the origin  (for positive shear) or away from the the origin  (negative shear),
so that a proton is either at potential minimum or maximum. }
\label{shearJet}
\end{figure}

\end{document}